\documentclass[amssymb, preprintnumbers, showpacs showkeys,aps,prl,superscriptaddress,twocolumn,nobibnotes]{revtex4-2}

\usepackage{color} 
\usepackage{xcolor} 
\usepackage{hyperref}
\usepackage{slashed}
\usepackage{graphicx}
\usepackage{amsmath}
\usepackage{latexsym}
\usepackage{epstopdf}
\usepackage{amsmath}
\usepackage{enumitem}
\usepackage{amssymb,amsmath}
\usepackage{multirow}
\usepackage{mathtools}
\usepackage{bbm}
\usepackage{bm}
\usepackage{amsfonts}
\usepackage{amssymb}
\usepackage{float}
\usepackage{calligra}
\usepackage{makecell}
\usepackage[normalem]{ulem}
\usepackage[USenglish,american]{babel}
\usepackage{subfigure}
\usepackage{orcidlink}

\expandafter\ifx\csname package@font\endcsname\relax\else
 \expandafter\expandafter
 \expandafter\usepackage
 \expandafter\expandafter
 \expandafter{\csname package@font\endcsname}%
\fi
\begin{document}

\title{Virtual defect mediated quantum melting in supersolids}

\author{Lazaros Tsaloukidis\orcidlink{0000-0003-1292-1037
}}%
\email{ltsalouk@pks.mpg.de}
\affiliation{Max Planck Institute for the Physics of Complex Systems, Nothnitzer Stra\ss e 38, 01187 Dresden, Germany}
\affiliation{W\"urzburg-Dresden Cluster of Excellence ct.qmat, 01187, Dresden, Germany}

\author{Matheus Grossklags\orcidlink{0009-0004-9656-0789}}%
\affiliation{Departamento de F\'\i sica, Universidade Federal de Santa Catarina, 88040-900 Florian\'opolis, Brazil}

\author{Luca Salasnich\orcidlink{0000-0003-0817-4753}}
\affiliation{Dipartimento di Fisica e Astronomia “Galileo Galilei, Università di Padova, INFN - Sezione di Padova, Via Marzolo 8, 35131 Padova, Italy}

\author{Alejandro Mendoza-Coto\orcidlink{0000-0001-5629-1095}}%
\email{alejandro.mendoza@ufsc.br}
\affiliation{Departamento de F\'\i sica, Universidade Federal de Santa Catarina, 88040-900 Florian\'opolis, Brazil}%
\begin{abstract}
    The current understanding of supersolid melting is largely based on mean-field theories predicting that such transitions arise from a structural energy minimization mechanism. Here, we show that soft-core quasi-one-dimensional supersolids stabilized by a shallow external lattice can instead melt at zero temperature into a superfluid phase with quasi-long-range positional order through the proliferation of virtual soliton--antisoliton pairs, leading to a Kosterlitz--Thouless transition. We derive the corresponding Renormalization Group equations from first principles and use them to determine the critical properties of the system. Remarkably, the proliferation of topological defects near criticality strongly renormalizes both the density modulation and the superfluidity behavior. The transition described is intrinsically a beyond mean-field effect, since the external lattice produces only a smooth crossover between weakly and strongly modulated states at the mean-field level.
\end{abstract}
\maketitle
\textit{Introduction.}---The physics of supersolids is currently a subject of intense research across fields ranging from atomic and molecular physics~\cite{cinti2010,cinti2014,zhang2019,matos2025,lima2025} to solid-state and condensed matter physics~\cite{joglekar2006,prestipino2018,pretko2018,kumar2019,bhowmick2021,mendoza2022,conti2023,zeng2026}. This broad interest has been greatly stimulated by the realization of supersolid phases in ultracold atomic systems under highly controlled conditions~\cite{leonard2017,li2017,tanzi2019,bottcher2019,natale2019}. A central theoretical question emerging from these developments concerns the phase diagram of supersolid systems and, in particular, the nature of the quantum supersolid to superfluid melting transition~\cite{zhang2019,smith2023,ripley2023,lima2025}. To a large extent, the current understanding of this transition at zero temperature is based on a mean-field picture in which melting is governed by an energy minimization process. Within this framework, the symmetries of the modulated phases involved, largely determines whether the transition is continuous or discontinuous, as expected from Ginzburg--Landau arguments~\cite{zhang2019,blakie2020,mendoza2021,mendoza2022,lima2025}.

This mean-field perspective, however, does not capture all mechanisms that can control the stability and melting of a supersolid. In many relevant regimes, quantum fluctuations play an essential role in shaping the properties of the system. A prominent example is the stabilization of self-bound dipolar droplets through the Lee--Huang--Yang correction~\cite{schutzhold2006,lima2011,lima2012}. Quantum fluctuations can also qualitatively modify the phase diagram of supersolid systems, for instance, by giving rise to second-order quantum melting transitions in planar geometries~\cite{zhang2019,ripley2023,lima2025}. Such scenario already shows how the nature of the supersolid melting in certain circumstances is not properly described once we neglect the role of quantum fluctuations. One such possibility, which remains largely unexplored in supersolid systems, is melting mediated by topological defects. Although real topological excitations are strongly suppressed at zero temperature because of their finite energy cost, it has been proposed that under certain conditions their virtual counterparts, i.e., dynamically created and annihilated defects having a finite extension in imaginary time direction, may still proliferate and drive the system to a quantum phase transition~\cite{lawler2004,sela2011,beekman2017}.

In this work, we show the existence of a defect-mediated supersolid--superfluid transition in a quasi-one-dimensional soft-core Bose gas, subjected to a shallow periodic potential. Starting from a microscopic description, we derive the low-energy elastic action for the deformation field of the supersolid and construct the perturbative Renormalization Group (RG) equations governing the effective rigidity and the external potential amplitude. We find that the external lattice stabilizes a true supersolid phase while allowing the creation of virtual soliton--antisoliton defect pairs. We show that the proliferation of these defects produce a Kosterlitz--Thouless (KT) melting transition between a supersolid phase with long-range positional order and a superfluid with quasi-long-range positional order at zero temperature. This topological phase transition has no mean-field counterpart, as in the present case, mean-field theory predicts only a smooth crossover between weakly and strongly modulated states.

\begin{figure*}[!t]
    \centering
    \includegraphics[width=\textwidth]{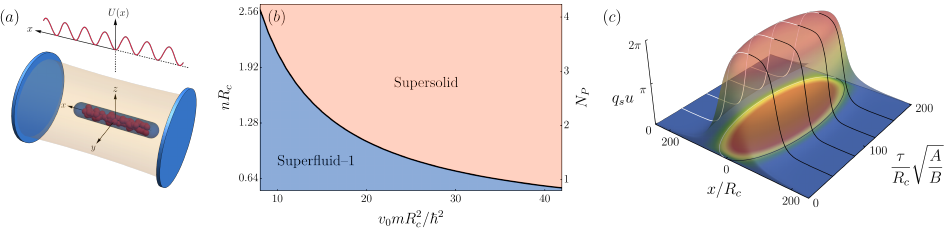}
    \caption{(a) Schematic experimental implementation of a soft-core bosonic gas on an optical lattice. The external potential stabilizes a long-range ordered supersolid phase. (b) Ground-state phase diagram for a system with effective pair potential $V_{\mathrm{eff}}(x)=v_{0}/\big(1+(x/R_{c})^{6}\big)$ on a commensurate substrate with amplitude $U_{0}mR^{2}_{c}/\hbar^{2}=0.01$ and wave vector $qR_{c}=3.93$, similar to the roton instability wave vector ($q_{0}R_{c}\approx3.93411$). Superfluid--$1$ phase corresponds to a modulated superfluid state with a single characteristic length scale and quasi-long-range positional order (c) Optimal soliton--antisoliton defect trajectory in dimensionless space-time coordinates for $nR_{c}=1.3$ and $v_{0}mR^{2}_{c}/\hbar^{2}=17$. The displayed trajectory produces the lowest action cost among all defect loops with internal area $\mathcal{A}\approx3\times10^{4}$.}
    \label{fig1}
\end{figure*}

\textit{Model and Method.}---As schematically shown in Fig.~\ref{fig1}(a), we consider a gas of $N$ soft-core bosons of mass $m$ strongly confined by a harmonic trap along the $y$ and $z$ directions. In the $x$-direction, the system has dimension $L$ and is subjected to an external monochromatic shallow lattice potential given by $U(x)=U_{0}\sin^{2}(qx/2)$. In the strong harmonic confinement regime, we can perform a ``dimensional reduction'', see Supp. Mat. for details~\cite{supplemental_material}, and focus on the study of the modulated phases developed by the system along the $x$-direction. The resulting $1$D effective Hamiltonian reads
\begin{equation}\label{effective_hamiltonian}
    \begin{split}
        \mathcal{H}[\psi,\psi^{\ast}]=&\int dx\bigg[\frac{\hbar^2}{2m}\lvert\nabla\psi(x)\rvert^{2}+U(x)\lvert\psi(x)\rvert^{2}\\
        &+\frac{1}{2}\int dx'V_{\mathrm{eff}}(x-x')\lvert\psi(x)\rvert^{2}\lvert\psi(x')\rvert^{2}\bigg],
    \end{split}
\end{equation}
where $\psi(x)$ is the bosonic complex field which satisfies the normalization condition $\int_{0}^{L}dx\lvert\psi(x)\rvert^{2}=N$. The first term in Eq.~\eqref{effective_hamiltonian} corresponds to the kinetic energy of the system, while the second contribution yields the potential energy due to the external optical lattice. Lastly, the non-local contribution represents the effective interaction produced by the soft-core two-body potential, modeled in this work as $V_{\mathrm{eff}}(x)=v_{0}/\big(1+(x/R_{c})^{6}\big)$, where $R_{c}$ and $v_{0}$ characterizes the range and overlap potential energy of the effective $1$D potential, respectively. The exact form of the model considered for $V_{\mathrm{eff}}(x)$ is not essential to this work since any monotonic soft-core interaction whose Fourier transform includes a negative sector displays a clustering behavior for large enough densities~\cite{cinti2014,shen2026}.

In order to study the phase diagram, we first determine the mean-field ground-state configuration $\psi_{0}(x)$, obtained from the minimization of the energy per particle functional $\mathcal{H}[\psi,\psi^{\ast}]/N$. This minimization is carried out by means of a spectral variational method that expands the wavefunction into a discrete Fourier basis and takes the Fourier amplitudes and wave vectors as variational parameters. This method has been successfully employed in the study of supersolids~\cite{zhang2019,prestipino2018,mendoza2021,lima2025}, quantum quasicrystals~\cite{mendoza2022,zampronio2024,grossklags2025,mendoza2026} and quasicrystalline Bose glass phases~\cite{grossklags2024}. Therefore, we consider the following ansatz for the ground-state wavefunction $\psi_{0}(x)=\sqrt{n}\sum_{j}c_{j}\exp(ijk_{0}x)/(2\Xi)$, where $c_{j}$ and $jk_{0}$, corresponds to the Fourier amplitudes and wave vectors of our expansion~\footnote{To guarantee a real valued wavefunction, we should impose the constraint $c_{j}=c_{-j}$ in the wavefunction ansatz.}, while $n=N/L$ stands for the linear density of particles and the parameter $\Xi=(\sum_{j}c^{2}_{j}/4)^{1/2}$ ensures the normalization condition of the wavefunction.

To access the properties of the system beyond mean-field, we consider its grand-canonical partition function $\mathcal{Z}=\int D\psi D\psi^{\ast}\exp(-\mathcal{S}[\psi,\psi^{\ast}])$, with action $\mathcal{S}[\psi,\psi^{\ast}]=\int d\tau dx~\psi^{\ast}(\hbar\partial_{\tau}-\mu)\psi+\int d\tau~\mathcal{H}[\psi,\psi^{\ast}]$. We then expand around the mean-field ground-state configuration $\psi_{0}(x)$ minimizing $\mathcal{H}[\psi,\psi^{\ast}]$ and $\mathcal{S}[\psi,\psi^{\ast}]$, allowing low-energy excited configurations with perturbed condensate and modulation phases, as well as in their respective conjugate density fields; see Supp. Mat.~\cite{supplemental_material}. This procedure yields an effective action $\mathcal{S}_\mathrm{eff}[u]$ for the supersolid deformation field $u(x,\tau)$ along the $x$-direction, while retaining its connection to the microscopic Hamiltonian. Similar constructions have been developed for supersolids~\cite{yoo2010,platt2024,poli2024,cook2026} and superquasicrystals~\cite{mendoza2025,mendoza2026}, following different routes but yielding consistent results. Here, it takes the form of a non-local quantum sine-Gordon model,
\begin{equation}\label{reduced_action2}
    \begin{split}
        &\mathcal{S}_{\mathrm{eff}}[u]=\frac{1}{2}\int\frac{dk d\omega}{(2\pi)^{2}}\bigg\{(Ak^{2}+B\omega^{2})+\frac{Ck^{2}\omega^{2}}{(Dk^{2}+E\omega^{2})}\bigg\}\\
        &\times\hat{u}(k,\omega)\hat{u}(-k,-\omega)
         -\int d\tau dx\kappa_{u}\cos(qu(x,\tau)),
    \end{split}
\end{equation}
where $\hat{u}(k,\omega)=\int dxd\tau\exp(-i(kx-\omega\tau))u(x,\tau)$. Moreover, the couplings $A$, $B$, $C$, $D$, $E$ and $\kappa_{u}$ are completely determined from the ground-state properties, for details see Supp. Mat.~\cite{supplemental_material}. After a space and time symmetrization of the action, standard methods in the study of the quantum sine-Gordon model~\cite{ohta1978,herbut2007} allow us to reach the perturbative RG equations for the effective stiffness and fugacity of the system
\begin{equation}\label{rg_flow_equations}
    \begin{split}
        \frac{da}{dl}&=\pi^{3}k_{u}(l),\\
        \frac{dk_{u}}{dl}&=\left(2-\frac{\pi}{a(l)}\frac{\sqrt{d}+\sqrt{e}}{\big((\sqrt{d}+\sqrt{e})^{2}+c/a(l)\big)^{1/2}}\right)k_{u}(l),
    \end{split}
\end{equation}
where $a(0)=\sqrt{AB}(2\pi/q)^{2}$, $c=\sqrt{AB}(2\pi/q)^{2}C$, $d=BD$, $e=EA$ and $k_{u}=\sqrt{B/A}(2\pi/q)^{2}\kappa_{u}$. Here, $a(l)$ represents the running stiffness of the symmetrized quadratic momentum-frequency interaction proportional to $(k'^{2}+\omega'^{2})$. Numerical integration of the RG equations reveals two distinct low energy regimes. In the first, $k_{u}(+\infty)=0$ while $a(+\infty)>0$, indicating that the locking term in the action $S_{\mathrm{eff}}[u]$ becomes irrelevant at long wavelengths. The displacement field then remains rough, corresponding to a modulated superfluid with quasi-long-range positional order. In the second, $k_{u}(+\infty)=+\infty$ and $a(+\infty)=+\infty$, leading to a displacement field effectively locked to a minimum of the periodic potential corresponding to a pinned supersolid with long-range positional order. The superfluid response remains finite in both regimes, hence the transition is controlled by the loss of long-range positional order. Consistently with known results for the quantum sine-Gordon limit $(C=0)$~\cite{amit1980}, the obtained model displays a positional correlation of the form $\langle\exp[iq(u(x,\tau)-u(0,0))]\rangle$ that approaches a non-zero constant at large space-time separations in the supersolid phase, whereas in the modulated--superfluid phase it decays algebraically as $\langle\exp[iq(u(x,\tau)-u(0,0))]\rangle\sim1/(x^{2}+A\tau^{2}/B)^{\eta/2}$, with $\eta\rightarrow4$ as we approach the phase boundary from the superfluid side.

\textit{Defect-mediated supersolid melting.}---As can be observed, in Fig.~\ref{fig1}(b) we present the $nR_{c}$ versus $v_{0}mR^{2}_{c}/\hbar^{2}$ phase diagram determined from our RG analysis for a shallow external lattice with amplitude $U_{0}mR^{2}_{c}/\hbar^{2}=0.01$ and wave vector $qR_{c}=3.93$, close to the roton wave vector of the free system $(q_{0}R_{c}\approx3.93411)$. The solid black line marks a continuous virtual defect-mediated transition between a locally modulated superfluid with quasi-long-range positional order (Superfluid--$1$ in Fig.~\ref{fig1}(b)) and a supersolid phase displaying long-range positional order (Supersolid region in Fig.~\ref{fig1}(b)). This transition is absent at the mean-field level, where the external lattice explicitly breaks translational symmetry and only a crossover between weakly and strongly modulated states occurs.

The RG flow associated to the non-local sine-Gordon model obtained, $S_{\mathrm{eff}}[u]$, displays a critical behavior consistent with the KT universality class, see End Matter for details. It is well understood that the KT transition of the quantum sine-Gordon model corresponds to a roughening transition of the field $u(x,\tau)$ in which the field evolves from a smooth, locked configuration, with $u(x,\tau)$ fluctuating weakly around a single minimum, to a rough regime where finite domains associated with different minima, $q_{s}u(x,\tau)=2\pi j$ with $j\in\mathbb{Z}$, are excited~\cite{flamino2020}. For our one-dimensional system, the excitation of two-dimensional domains in the $x$--$\tau$ plane corresponds to a dynamical process in which a soliton--antisoliton pair is nucleated at a given $\tau$, propagates over a finite imaginary-time interval and eventually annihilates, thereby forming a closed loop in spacetime as shown in Fig.~\ref{fig1}(c). The action cost of such excitations increases with the length of its boundary, consequently, isolated defects extending throughout the infinite imaginary-time direction are forbidden at zero temperature, whereas finite virtual soliton--antisoliton loops remain the leading mechanism to induce disorder in the periodic order of the supersolid phase. To illustrate quantitatively this microscopic mechanism, we determine the lowest action loop configuration in the $x$--$\tau$ plane subject to the constraint of having a fixed enclosed area, see Supp. Matt. for details~\cite{supplemental_material}. An example of the resulting field configuration, evaluated at a point close to the supersolid phase boundary ($nR_{c}=1.3$ and $v_{0}mR^{2}_{c}/\hbar^{2}=17$), is shown in Fig.~\ref{fig1}(c), where the white (black) contours represent the propagating soliton (antisoliton) along imaginary time.

\begin{figure}[!t]
    \centering
    \includegraphics[width=0.48\textwidth]{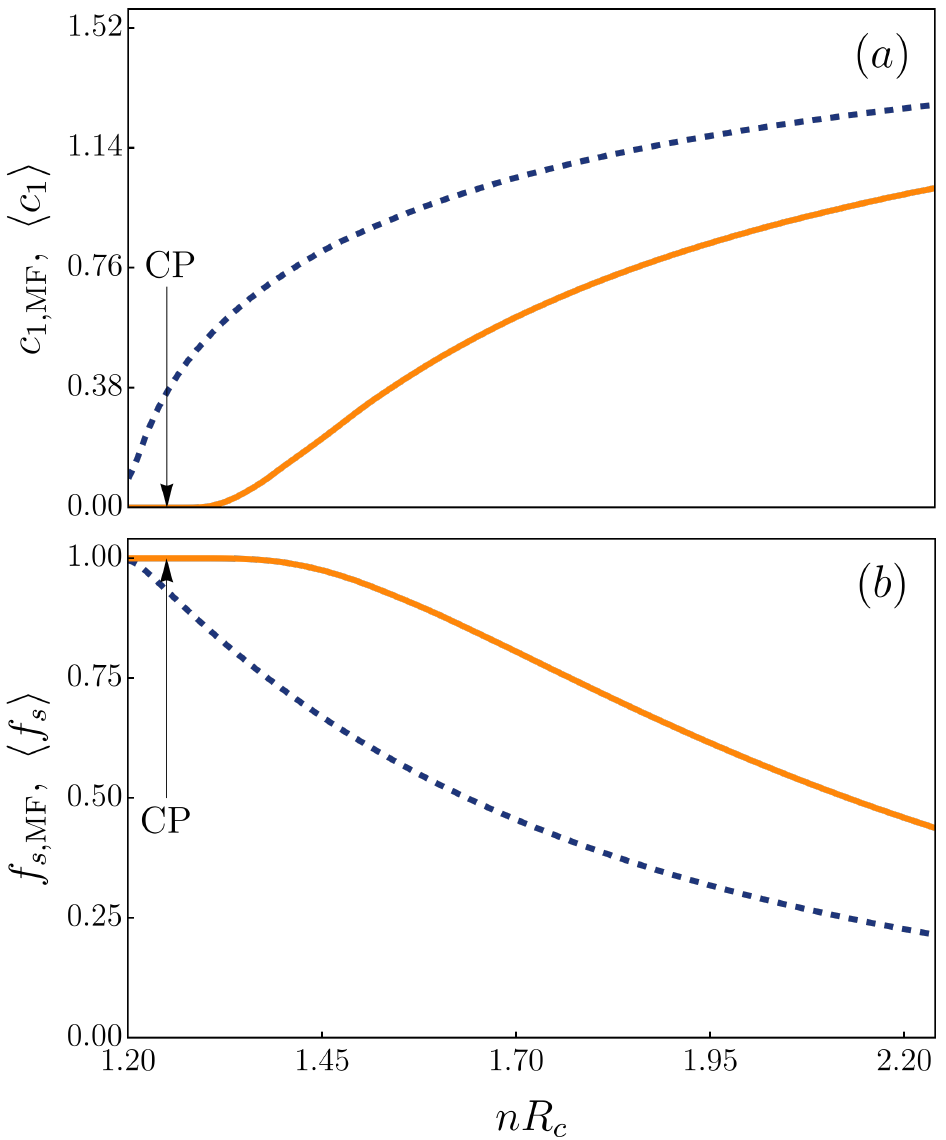}
    \caption{(a) Leading supersolid amplitude $c_{1}$ calculated within the mean-field approximation (dashed blue curve) and corrected according to RG calculations (solid orange curve) as a function of dimensionless density $nR_{c}$. (b) Superfluid fraction $f_{s}$ associated to the mean-field (dashed blue curve) and renormalized (solid orange curve) density profile as a function of dimensionless density $nR_{c}$. For both figures, we consider $v_{0}mR^{2}_{c}/\hbar^{2}=17$ and $U_{0}mR^{2}_{c}/\hbar^{2}=0.01$, while the position of the critical point is signaled with CP.}
    \label{fig2}
\end{figure}

\textit{Consequences of the topological phase transition.}---The occurrence of a KT superfluid-to-supersolid transition has direct implications for the behavior of observables near the phase boundary. First, the correlation length follows the well known KT critical behavior of the form $\xi(\tau)\propto\exp(b/\sqrt{\tau})$, where $\tau$ measures the distance from the critical line. As expected, this singular behavior strongly impacts the spatial properties of the supersolid density modulation. To quantify this effect, we expand the mean-field density profile in Fourier components and include positional fluctuations through the displacement field $u(x,\tau)$. Hence, the average density pattern can be written as $\langle\rho(x,\tau)\rangle=\sum_{j\geq0}\langle c_{j}\rangle\cos(jqx)$, with $\langle c_{j}\rangle=c_{j,\mathrm{MF}}\langle\exp[ijqu(x,\tau)]\rangle$, where the MF label indicates evaluation within the mean-field approximation. At this point, the expectation value $\langle\exp(ijqu)\rangle$ is estimated within the perturbative RG treatment, as discussed the in End Matter section. Close to the critical boundary, we obtain that $\langle c_{j}\rangle=c_{j,\mathrm{MF}}\exp[-j^{2}G/\sqrt{n-n_{c}}]$, where $G$ is a non-universal constant. Although the bare mean-field amplitudes evolve smoothly across the transition region, positional fluctuations produce an essential suppression of the density modulation as the critical density $n_{c}$ is approached.

\begin{figure}[!t]
    \centering
    \includegraphics[width=0.48\textwidth]{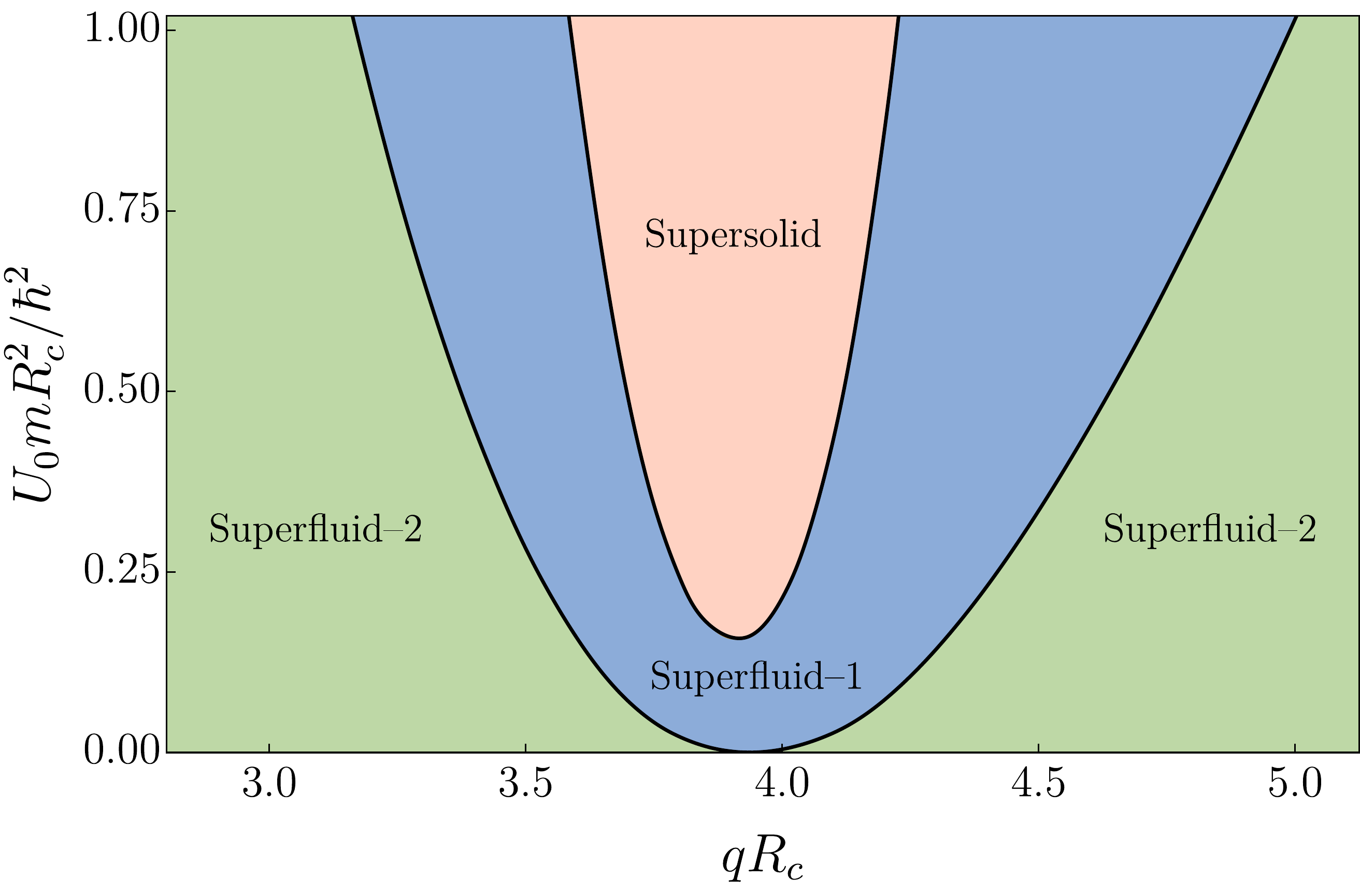}
    \caption{Supersolid phase diagram for $nR_{c}=0.7$ and $v_{0}mR^{2}_{c}/\hbar^{2}=30$. The Superfluid--$1(2)$ phase corresponds to global superfluid states presenting local modulations in the density with one (two) characteristic wave vectors.}
    \label{fig3}
\end{figure}

A similar renormalization effect can be observed in the superfluid response within the critical region. At the mean-field level, the one-dimensional superfluid fraction can be determined via minimization of the static phase gradient energy $\Delta E[\theta]=(\hbar^{2}/2m)\int dx\rho(x)(\partial_{x}\theta)^{2}$, under a fixed phase twist to yield the superfluid fraction~\cite{platt2024} $f_{s,\mathrm{MF}}=\left(\int dx\rho(x)/L\times\int dx\rho(x)^{-1}/L\right)^{-1}$. To incorporate the effects of quantum fluctuations, including soliton--antisoliton defects, we can integrate the displacement fields while retaining the condensate phase. In this scenario, the first cumulant of the static phase gradient coupling replaces $\rho(x)$ by the equilibrium average $\langle\rho(x)\rangle_{0}$. Thus, an application of the same phase twist minimization leads to an equivalent expression for the superfluid fraction, see End Matter for details. Considering the leading order correction to $c_{1}$, the renormalized superfluid fraction reduces to $\langle f_{s}\rangle\approx1-(c_{1,\mathrm{MF}}/c_{0,\mathrm{MF}})^{2}\exp[-2G/\sqrt{n-n_{c}}]$. In Figs.~\ref{fig2}(a)-(b), we compare mean-field and RG-corrected predictions for $c_{1}$ and $f_{s}$ taking $nR_{c}$ as running parameter. A pronounced departure from the smooth mean-field crossover emerges close to the phase boundary, providing a clear signature of the underlying KT criticality. This strong renormalization originates from the increasing contribution of topological-defect fluctuations upon approaching the transition. Although quantitative details may be modified beyond the perturbative treatment, the essential KT-like renormalization of the density modulation and superfluid response is expected to persist in a non-perturbative description.

\textit{Melting induced by frustration.}---In the system considered, two competing ordering tendencies coexist. On the one hand, the combined effects of kinetic energy and two-body interactions favor a density modulation with a wave vector close to that of the roton minimum $(q_{\mathrm{rot}})$, while on the other hand, the external lattice favors a modulation with wave vector $q$. For $q$ sufficiently close to $q_{\mathrm{rot}}$ and strong enough lattice amplitudes $U_{0}$, the mean-field ground-state can still adjust its modulation wave vector $k_{0}$ to that of the external lattice $q$. However, such locking comes with an energetic cost and progressively weakens the mean-field supersolid modulation as $\vert q-q_{\mathrm{rot}}\vert$ increases. Hence, when the frustration between $q$ and $q_\mathrm{rot}$ is strong enough, the competition between the two different ordering tendencies in the system produces a mean-field phase transition from a crystalline to an aperiodic phase. The location of such phase transition can be obtained including in our variational study of the energy per particle $\mathcal{H}[\psi,\psi^{\ast}]/N$, the possibility of a quasiperiodic $\psi_{0}(x)$, generated from a Fourier expansion with a two wave vector basis~\cite{mendoza2022,grossklags2024,mendoza2026}. 

Beyond this mean-field picture, which in essence captures the local structure of the density pattern, we now employ the RG formalism developed to locate the long range positionally ordered supersolid phase varying the properties of the external lattice. The obtained results are shown in Fig.~\ref{fig3} where the $qR_{c}$ versus $U_{0}mR^{2}_{c}/\hbar^{2}$ phase diagram is displayed for $nR_{c}=0.7$ and $v_{0}mR^{2}_{c}/\hbar^{2}=30$. As can be observed, we identify three different phases, a RG-stable supersolid and two superfluid phases locally modulated with one (Superfluid--$1$) and two (Superfluid--$2$) incommensurate characteristic wave vectors. Although the occurrence of the supersolid dome around $q=q_\mathrm{rot}$ was already expected, the rapidly growing behavior of its boundary with $U_{0}$ shows how effective the competition between the two ordering tendencies in the system is in suppressing the supersolid phase and also shows how accurately $q$ must be tuned in order to access the defect mediated supersolid--superfluid transition.

\textit{Conclusions.}---We have shown that a suitably tuned external periodic potential can stabilize a long-range-ordered supersolid phase in quasi-one-dimensional soft-core interacting bosonic gases. Remarkably, this ordered state can melt at zero temperature through the proliferation of virtual topological defect pairs, destroying long-range positional order through a mechanism absent at the mean-field level. In contrast to previously studied scenarios in which defects are energetically allowed within ground-state configurations~\cite{radzi2020,nguyen2020,kumar2019,beekman2017}, the defects considered here are intrinsically dynamical, occurring only with finite extent in imaginary time as part of quantum fluctuations. Although such a scenario has been proposed previously~\cite{lawler2004,sela2011,beekman2017}, here we identify a microscopic model in which defects extending indefinitely in imaginary time are suppressed, while virtual defect--antidefect configurations proliferate and drive the transition. The resulting supersolid--superfluid transition mediated by virtual defects thus provides a distinct mechanism for supersolid melting. More broadly, the fact that low-energy virtual defects can qualitatively modify the mean-field transition raises the question of whether virtual dislocations may similarly affect supersolid melting in two and three dimensions, particularly in regimes where mean-field theory predicts a continuous or weakly first-order transition~\cite{zhang2019,ripley2023,lima2025}.

Our results also establish experimentally accessible signatures of this process. Approaching the transition, the proliferation of virtual topological defects produces a strong KT-like renormalization of both the Fourier amplitudes of the density modulation and the superfluid response, in sharp contrast with the smooth crossover predicted at the mean-field level. This behavior can be tested experimentally or numerically, providing a direct probe of the defect-mediated melting mechanism.

\textit{Acknowledgments.}---A.M.C. acknowledges financial support from the Brazilian National Council for Scientific
and Technological Development (CNPq), under Grant
No. 311122/2026-4, and thanks the Max Planck Insti-
tute for the Physics of Complex Systems (MPIPKS) for
its financial support and hospitality. L.S. is partially supported by the Project “Frontiere Quantistiche” (Dipartimenti di Eccellenza) of the Italian Ministry of University and Research and by “Iniziativa Specifica Quantum” of INFN. L.T. was supported in part by the Deutsche Forschungsgemeinschaft under cluster of excellence ct.qmat (EXC 2147, Project-ID No. 390858490).
\bibliography{ref.bib}

@article{mendoza2022,
    title = {Exploring quantum quasicrystal patterns: A variational study},
    author = {Mendoza-Coto, A. and Turcati, R. and Zampronio, V. and D\'{\i}az-M\'endez, R. and Macr\`{\i}, T. and Cinti, F.},
    journal = {Phys. Rev. B},
    volume = {105},
    issue = {13},
    pages = {134521},
    numpages = {11},
    year = {2022},
    month = {Apr},
    publisher = {American Physical Society},
    doi = {10.1103/PhysRevB.105.134521},
    url = {https://link.aps.org/doi/10.1103/PhysRevB.105.134521}
}

@article{platt2024,
    title = {Sound waves and fluctuations in one-dimensional supersolids},
    author = {Platt, L. M. and Baillie, D. and Blakie, P. B.},
    journal = {Phys. Rev. A},
    volume = {110},
    issue = {2},
    pages = {023320},
    numpages = {14},
    year = {2024},
    month = {Aug},
    publisher = {American Physical Society},
    doi = {10.1103/PhysRevA.110.023320},
    url = {https://link.aps.org/doi/10.1103/PhysRevA.110.023320}
}

@article{yoo2010,
    title = {Hydrodynamic theory of supersolids: Variational principle, effective Lagrangian, and density-density correlation function},
    author = {Yoo, C.-D. and Dorsey, Alan T.},
    journal = {Phys. Rev. B},
    volume = {81},
    issue = {13},
    pages = {134518},
    numpages = {12},
    year = {2010},
    month = {Apr},
    publisher = {American Physical Society},
    doi = {10.1103/PhysRevB.81.134518},
    url = {https://link.aps.org/doi/10.1103/PhysRevB.81.134518}
}

@article{leonard2017,
    author = {L{\'e}onard, Julian and Morales, Andrea and Zupancic, Philip and Esslinger, Tilman and Donner, Tobias},
    title = {Supersolid formation in a quantum gas breaking a continuous translational symmetry},
    journal = {Nature},
    year = {2017},
    month = {Mar},
    day = {01},
    volume = {543},
    number = {7643},
    pages = {87-90},
    issn = {1476-4687},
    doi = {10.1038/nature21067},
    url = {https://doi.org/10.1038/nature21067}
}

@article{li2017,
    author = {Li, Jun-Ru and Lee, Jeongwon and Huang, Wujie and Burchesky, Sean and Shteynas, Boris and Top, Furkan {\c{C}}a{\u{g}}r{\i} and Jamison, Alan O. and Ketterle, Wolfgang},
    title = {A stripe phase with supersolid properties in spin--orbit-coupled Bose--Einstein condensates},
    journal = {Nature},
    year = {2017},
    month = {Mar},
    day = {01},
    volume = {543},
    number = {7643},
    pages = {91-94},
    issn = {1476-4687},
    doi = {10.1038/nature21431},
    url = {https://doi.org/10.1038/nature21431}
}

@article{tanzi2019,
    title = {Observation of a Dipolar Quantum Gas with Metastable Supersolid Properties},
    author = {Tanzi, L. and Lucioni, E. and Fam\`a, F. and Catani, J. and Fioretti, A. and Gabbanini, C. and Bisset, R. N. and Santos, L. and Modugno, G.},
    journal = {Phys. Rev. Lett.},
    volume = {122},
    issue = {13},
    pages = {130405},
    numpages = {6},
    year = {2019},
    month = {Apr},
    publisher = {American Physical Society},
    doi = {10.1103/PhysRevLett.122.130405},
    url = {https://link.aps.org/doi/10.1103/PhysRevLett.122.130405}
}

@article{bottcher2019,
    title = {Transient Supersolid Properties in an Array of Dipolar Quantum Droplets},
    author = {B\"ottcher, Fabian and Schmidt, Jan-Niklas and Wenzel, Matthias and Hertkorn, Jens and Guo, Mingyang and Langen, Tim and Pfau, Tilman},
    journal = {Phys. Rev. X},
    volume = {9},
    issue = {1},
    pages = {011051},
    numpages = {7},
    year = {2019},
    month = {Mar},
    publisher = {American Physical Society},
    doi = {10.1103/PhysRevX.9.011051},
    url = {https://link.aps.org/doi/10.1103/PhysRevX.9.011051}
}

@article{natale2019,
    title = {Excitation Spectrum of a Trapped Dipolar Supersolid and Its Experimental Evidence},
    author = {Natale, G. and van Bijnen, R. M. W. and Patscheider, A. and Petter, D. and Mark, M. J. and Chomaz, L. and Ferlaino, F.},
    journal = {Phys. Rev. Lett.},
    volume = {123},
    issue = {5},
    pages = {050402},
    numpages = {6},
    year = {2019},
    month = {Aug},
    publisher = {American Physical Society},
    doi = {10.1103/PhysRevLett.123.050402},
    url = {https://link.aps.org/doi/10.1103/PhysRevLett.123.050402}
}

@article{prestipino2018,
    title = {Freezing of soft-core bosons at zero temperature: A variational theory},
    author = {Prestipino, Santi and Sergi, Alessandro and Bruno, Ezio},
    journal = {Phys. Rev. B},
    volume = {98},
    issue = {10},
    pages = {104104},
    numpages = {16},
    year = {2018},
    month = {Sep},
    publisher = {American Physical Society},
    doi = {10.1103/PhysRevB.98.104104},
    url = {https://link.aps.org/doi/10.1103/PhysRevB.98.104104}
}

@article{conti2023,
    title = {Chester Supersolid of Spatially Indirect Excitons in Double-Layer Semiconductor Heterostructures},
    author = {Conti, Sara and Perali, Andrea and Hamilton, Alexander R. and Milo\ifmmode \check{s}\else \v{s}\fi{}evi\ifmmode \acute{c}\else \'{c}\fi{}, Milorad V. and Peeters, Fran\ifmmode \mbox{\c{c}}\else \c{c}\fi{}ois M. and Neilson, David},
    journal = {Phys. Rev. Lett.},
    volume = {130},
    issue = {5},
    pages = {057001},
    numpages = {6},
    year = {2023},
    month = {Feb},
    publisher = {American Physical Society},
    doi = {10.1103/PhysRevLett.130.057001},
    url = {https://link.aps.org/doi/10.1103/PhysRevLett.130.057001}
}

@article{joglekar2006,
    title = {Wigner supersolid of excitons in electron-hole bilayers},
    author = {Joglekar, Yogesh N. and Balatsky, Alexander V. and Das Sarma, S.},
    journal = {Phys. Rev. B},
    volume = {74},
    issue = {23},
    pages = {233302},
    numpages = {4},
    year = {2006},
    month = {Dec},
    publisher = {American Physical Society},
    doi = {10.1103/PhysRevB.74.233302},
    url = {https://link.aps.org/doi/10.1103/PhysRevB.74.233302}
}

@book{Kardar2007,
    author    = {Kardar, Mehran},
    title     = {Statistical Physics of Fields},
    publisher = {Cambridge University Press},
    year      = {2007},
    isbn      = {9780521873413},
    doi       = {10.1017/CBO9780511815881}
}

@article{zeng2026,
    author = {Zeng, Yihang and Sun, Dihao and Zhang, Naiyuan J. and Nguyen, Ron Q. and Shi, Qianhui and Okounkova, A. and Watanabe, K. and Taniguchi, T. and Hone, J. and Dean, C. R. and Li, J. I. A.},
    title = {Observation of a superfluid-to-insulator transition of bilayer excitons},
    journal = {Nature},
    year = {2026},
    month = {Feb},
    day = {01},
    volume = {650},
    number = {8100},
    pages = {86-92},
    issn = {1476-4687},
    doi = {10.1038/s41586-025-09986-w},
    url = {https://doi.org/10.1038/s41586-025-09986-w}
}

@article{cinti2014,
    author = {Cinti, F. and Macr{\`i}, T. and Lechner, W. and Pupillo, G. and Pohl, T.},
    title = {Defect-induced supersolidity with soft-core bosons},
    journal = {Nature Communications},
    year = {2014},
    month = {Feb},
    day = {04},
    volume = {5},
    number = {1},
    pages = {3235},
    issn = {2041-1723},
    doi = {10.1038/ncomms4235},
    url = {https://doi.org/10.1038/ncomms4235}
}

@article{cinti2010,
    title = {Supersolid Droplet Crystal in a Dipole-Blockaded Gas},
    author = {Cinti, F. and Jain, P. and Boninsegni, M. and Micheli, A. and Zoller, P. and Pupillo, G.},
    journal = {Phys. Rev. Lett.},
    volume = {105},
    issue = {13},
    pages = {135301},
    numpages = {4},
    year = {2010},
    month = {Sep},
    publisher = {American Physical Society},
    doi = {10.1103/PhysRevLett.105.135301},
    url = {https://link.aps.org/doi/10.1103/PhysRevLett.105.135301}
}

@article{poli2024,
    title = {Excitations of a two-dimensional supersolid},
    author = {Poli, E. and Baillie, D. and Ferlaino, F. and Blakie, P. B.},
    journal = {Phys. Rev. A},
    volume = {110},
    issue = {5},
    pages = {053301},
    numpages = {10},
    year = {2024},
    month = {Nov},
    publisher = {American Physical Society},
    doi = {10.1103/PhysRevA.110.053301},
    url = {https://link.aps.org/doi/10.1103/PhysRevA.110.053301}
}

@article{grossklags2025,
    title = {Engineering interaction potentials for stabilizing quantum quasicrystal phases},
    author = {Grossklags, Matheus and Lima, Daniel and Zampronio, Vinicius and Cinti, Fabio and Mendoza-Coto, Alejandro},
    journal = {Phys. Rev. B},
    volume = {112},
    issue = {22},
    pages = {224107},
    numpages = {15},
    year = {2025},
    month = {Dec},
    publisher = {American Physical Society},
    doi = {10.1103/vnt6-9g1l},
    url = {https://link.aps.org/doi/10.1103/vnt6-9g1l}
}

@misc{shen2026,
    title = {Phase diagram of one-dimensional bosons with Rydberg-dressed soft-core interactions}, 
    author = {Shengjie Yu and Laurent Sanchez-Palencia},
    year = {2026},
    eprint = {2607.13623},
    archivePrefix = {arXiv},
    primaryClass = {cond-mat.quant-gas},
    url = {https://arxiv.org/abs/2607.13623}, 
}

@article{grossklags2024,
    title = {Self-induced Bose glass phase in quantum quasicrystals},
    journal = {Results in Physics},
    volume = {65},
    pages = {107991},
    year = {2024},
    issn = {2211-3797},
    doi = {https://doi.org/10.1016/j.rinp.2024.107991},
    url = {https://www.sciencedirect.com/science/article/pii/S2211379724006764},
    author = {M. Grossklags and M. Ciardi and V. Zampronio and F. Cinti and A. Mendoza-Coto}
}

@article{zampronio2024,
    title = {Exploring Quantum Phases of Dipolar Gases through Quasicrystalline Confinement},
    author = {Zampronio, Vinicius and Mendoza-Coto, Alejandro and Macr\`{\i}, Tommaso and Cinti, Fabio},
    journal = {Phys. Rev. Lett.},
    volume = {133},
    issue = {19},
    pages = {196001},
    numpages = {7},
    year = {2024},
    month = {Nov},
    publisher = {American Physical Society},
    doi = {10.1103/PhysRevLett.133.196001},
    url = {https://link.aps.org/doi/10.1103/PhysRevLett.133.196001}
}

@misc{mendoza2026,
    title = {Coexistence of long- and quasi-long range spatial order in 1D quantum quasicrystals}, 
    author = {A. Mendoza-Coto and M. Grossklags and J. Stefaniak and T. Donner and F. Piazza},
    year = {2026},
    eprint = {2607.17334},
    archivePrefix = {arXiv},
    primaryClass = {cond-mat.quant-gas},
    url = {https://arxiv.org/abs/2607.17334}, 
}

@article{cook2026,
    title = {Excitations and anisotropic sound in planar dipolar supersolids with tilted dipoles},
    author = {Cook, Reuben and Lee, Au-Chen and Blakie, P. Blair},
    journal = {Phys. Rev. A},
    volume = {113},
    issue = {6},
    pages = {063308},
    numpages = {14},
    year = {2026},
    month = {Jun},
    publisher = {American Physical Society},
    doi = {10.1103/32c1-vrbl},
    url = {https://link.aps.org/doi/10.1103/32c1-vrbl}
}

@article{zhang2019,
    title = {Supersolidity around a Critical Point in Dipolar Bose-Einstein Condensates},
    author = {Zhang, Yong-Chang and Maucher, Fabian and Pohl, Thomas},
    journal = {Phys. Rev. Lett.},
    volume = {123},
    issue = {1},
    pages = {015301},
    numpages = {6},
    year = {2019},
    month = {Jul},
    publisher = {American Physical Society},
    doi = {10.1103/PhysRevLett.123.015301},
    url = {https://link.aps.org/doi/10.1103/PhysRevLett.123.015301}
}

@article{matos2025,
    title = {Transitional Supersolidity in Ion Doped Helium Droplets},
    author = {Matos, Juan Carlos Acosta and Giannakeas, P. and Ciardi, Matteo and Pohl, Thomas and Rost, Jan M.},
    journal = {Phys. Rev. Lett.},
    volume = {134},
    issue = {18},
    pages = {186001},
    numpages = {7},
    year = {2025},
    month = {May},
    publisher = {American Physical Society},
    doi = {10.1103/PhysRevLett.134.186001},
    url = {https://link.aps.org/doi/10.1103/PhysRevLett.134.186001}
}

@article{flamino2020,
    title = {Lattice sine-Gordon model},
    author = {Flamino, James and Giedt, Joel},
    journal = {Phys. Rev. D},
    volume = {101},
    issue = {7},
    pages = {074503},
    numpages = {18},
    year = {2020},
    month = {Apr},
    publisher = {American Physical Society},
    doi = {10.1103/PhysRevD.101.074503},
    url = {https://link.aps.org/doi/10.1103/PhysRevD.101.074503}
}

@article{ohta1978,
    author = {Ohta, Takao},
    title = {Phase Transition in 2-D Sine-Gordon System},
    journal = {Progress of Theoretical Physics},
    volume = {60},
    number = {4},
    pages = {968-974},
    year = {1978},
    month = {10},
    issn = {0033-068X},
    doi = {10.1143/PTP.60.968},
    url = {https://doi.org/10.1143/PTP.60.968}
}

@book{herbut2007,
    author = {Herbut, Igor},
    title = {A Modern Approach to Critical Phenomena},
    publisher = {Cambridge University Press},
    address = {Cambridge},
    year = {2007},
    isbn = {978-0-521-84198-6}
}

@article{nguyen2020,
    title = {Fracton-elasticity duality of two-dimensional superfluid vortex crystals: defect interactions and quantum melting},
    pages = {076},
    author = {Nguyen, Dung and Gromov, Andrey and Moroz, Sergej},
    journal = {SciPost Phys.},
    volume = {9},
	year = {2020},
	publisher = {SciPost},
	doi = {10.21468/SciPostPhys.9.5.076},
	url = {https://scipost.org/10.21468/SciPostPhys.9.5.076}
}

@article{beekman2017,
    title = {Dual gauge field theory of quantum liquid crystals in two dimensions},
    journal = {Physics Reports},
    volume = {683},
    pages = {1-110},
    year = {2017},
    issn = {0370-1573},
    doi = {https://doi.org/10.1016/j.physrep.2017.03.004},
    url = {https://www.sciencedirect.com/science/article/pii/S0370157317300583},
    author = {Aron J. Beekman and Jaakko Nissinen and Kai Wu and Ke Liu and Robert-Jan Slager and Zohar Nussinov and Vladimir Cvetkovic and Jan Zaanen}
}

@article{sela2011,
    title = {Dislocation-mediated melting of one-dimensional Rydberg crystals},
    author = {Sela, Eran and Punk, Matthias and Garst, Markus},
    journal = {Phys. Rev. B},
    volume = {84},
    issue = {8},
    pages = {085434},
    numpages = {19},
    year = {2011},
    month = {Aug},
    publisher = {American Physical Society},
    doi = {10.1103/PhysRevB.84.085434},
    url = {https://link.aps.org/doi/10.1103/PhysRevB.84.085434}
}

@article{lima2025,
    title = {Supersolid dipolar phases in planar geometry: Effects of tilted polarization},
    author = {Lima, Daniel and Grossklags, Matheus and Zampronio, Vinicius and Cinti, Fabio and Mendoza-Coto, Alejandro},
    journal = {Phys. Rev. A},
    volume = {111},
    issue = {6},
    pages = {063311},
    numpages = {11},
    year = {2025},
    month = {Jun},
    publisher = {American Physical Society},
    doi = {10.1103/mlfn-114m},
    url = {https://link.aps.org/doi/10.1103/mlfn-114m}
}

@article{mendoza2025,
    title = {Low-Energy Excitations in Bosonic Quantum Quasicrystals},
    author = {Mendoza-Coto, A. and Bonifacio, M. and Piazza, F.},
    journal = {Phys. Rev. Lett.},
    volume = {134},
    issue = {13},
    pages = {136003},
    numpages = {7},
    year = {2025},
    month = {Apr},
    publisher = {American Physical Society},
    doi = {10.1103/PhysRevLett.134.136003},
    url = {https://link.aps.org/doi/10.1103/PhysRevLett.134.136003}
}

@article{kumar2019,
    title = {Symmetry-enforced fractonicity and two-dimensional quantum crystal melting},
    author = {Kumar, Ajesh and Potter, Andrew C.},
    journal = {Phys. Rev. B},
    volume = {100},
    issue = {4},
    pages = {045119},
    numpages = {13},
    year = {2019},
    month = {Jul},
    publisher = {American Physical Society},
    doi = {10.1103/PhysRevB.100.045119},
    url = {https://link.aps.org/doi/10.1103/PhysRevB.100.045119}
}

@article{radzi2020,
    title = {Quantum Smectic Gauge Theory},
    author = {Radzihovsky, Leo},
    journal = {Phys. Rev. Lett.},
    volume = {125},
    issue = {26},
    pages = {267601},
    numpages = {5},
    year = {2020},
    month = {Dec},
    publisher = {American Physical Society},
    doi = {10.1103/PhysRevLett.125.267601},
    url = {https://link.aps.org/doi/10.1103/PhysRevLett.125.267601}
}

@article{lawler2004,
    title = {Quantum Hall smectics, sliding symmetry, and the renormalization group},
    author = {Lawler, Michael J. and Fradkin, Eduardo},
    journal = {Phys. Rev. B},
    volume = {70},
    issue = {16},
    pages = {165310},
    numpages = {7},
    year = {2004},
    month = {Oct},
    publisher = {American Physical Society},
    doi = {10.1103/PhysRevB.70.165310},
    url = {https://link.aps.org/doi/10.1103/PhysRevB.70.165310}
}

@article{bhowmick2021,
    title = {Twisted superfluid and supersolid phases of triplons in bilayer honeycomb magnets},
    author = {Bhowmick, Dhiman and Chatterjee, Abhinava and Panigrahi, Prasanta K. and Sengupta, Pinaki},
    journal = {Phys. Rev. B},
    volume = {104},
    issue = {6},
    pages = {064447},
    numpages = {10},
    year = {2021},
    month = {Aug},
    publisher = {American Physical Society},
    doi = {10.1103/PhysRevB.104.064447},
    url = {https://link.aps.org/doi/10.1103/PhysRevB.104.064447}
}

@article{smith2023,
    title = {Supersolidity and crystallization of a dipolar Bose gas in an infinite tube},
    author = {Smith, Joseph C. and Baillie, D. and Blakie, P. B.},
    journal = {Phys. Rev. A},
    volume = {107},
    issue = {3},
    pages = {033301},
    numpages = {10},
    year = {2023},
    month = {Mar},
    publisher = {American Physical Society},
    doi = {10.1103/PhysRevA.107.033301},
    url = {https://link.aps.org/doi/10.1103/PhysRevA.107.033301}
}

@article{ripley2023,
    title = {Two-dimensional supersolidity in a planar dipolar Bose gas},
    author = {Ripley, B. T. E. and Baillie, D. and Blakie, P. B.},
    journal = {Phys. Rev. A},
    volume = {108},
    issue = {5},
    pages = {053321},
    numpages = {10},
    year = {2023},
    month = {Nov},
    publisher = {American Physical Society},
    doi = {10.1103/PhysRevA.108.053321},
    url = {https://link.aps.org/doi/10.1103/PhysRevA.108.053321}
}

@article{blakie2020,
    title = {Supersolidity in an elongated dipolar condensate},
    author = {Blakie, P. B. and Baillie, D. and Chomaz, L. and Ferlaino, F.},
    journal = {Phys. Rev. Res.},
    volume = {2},
    issue = {4},
    pages = {043318},
    numpages = {7},
    year = {2020},
    month = {Dec},
    publisher = {American Physical Society},
    doi = {10.1103/PhysRevResearch.2.043318},
    url = {https://link.aps.org/doi/10.1103/PhysRevResearch.2.043318}
}

@article{mendoza2021,
    title = {Ground-state phase diagram of ultrasoft bosons},
    author = {Mendoza-Coto, Alejandro and Caetano, Diogo de Souza and D\'{\i}az-M\'endez, Rogelio},
    journal = {Phys. Rev. A},
    volume = {104},
    issue = {1},
    pages = {013301},
    numpages = {7},
    year = {2021},
    month = {Jul},
    publisher = {American Physical Society},
    doi = {10.1103/PhysRevA.104.013301},
    url = {https://link.aps.org/doi/10.1103/PhysRevA.104.013301}
}

@article{schutzhold2006,
    author = {Sch\"{u}tzhold, Ralf and Uhlmann, Michael and Xu, Yan and Fischer, Uwe R.},
    title = {Mean-field expansion in Bose-Einstein condensates with finite-range interactions},
    journal = {International Journal of Modern Physics B},
    volume = {20},
    number = {24},
    pages = {3555-3565},
    year = {2006},
    doi = {10.1142/S0217979206035631},
    url = {https://doi.org/10.1142/S0217979206035631}
}

@article{lima2011,
    title = {Quantum fluctuations in dipolar Bose gases},
    author = {Lima, Aristeu R. P. and Pelster, Axel},
    journal = {Phys. Rev. A},
    volume = {84},
    issue = {4},
    pages = {041604(R)},
    numpages = {4},
    year = {2011},
    month = {Oct},
    publisher = {American Physical Society},
    doi = {10.1103/PhysRevA.84.041604},
    url = {https://link.aps.org/doi/10.1103/PhysRevA.84.041604}
}

@article{amit1980,
    doi = {10.1088/0305-4470/13/2/024},
    url = {https://doi.org/10.1088/0305-4470/13/2/024},
    year = {1980},
    month = {feb},
    publisher = {},
    volume = {13},
    number = {2},
    pages = {585},
    author = {D J Amit and Y Y Goldschmidt and S Grinstein},
    title = {Renormalisation group analysis of the phase transition in the 2D Coulomb gas, Sine-Gordon theory and XY-model},
    journal = {Journal of Physics A: Mathematical and General}
}

@article{lima2012,
    title = {Beyond mean-field low-lying excitations of dipolar Bose gases},
    author = {Lima, A. R. P. and Pelster, A.},
    journal = {Phys. Rev. A},
    volume = {86},
    issue = {6},
    pages = {063609},
    numpages = {16},
    year = {2012},
    month = {Dec},
    publisher = {American Physical Society},
    doi = {10.1103/PhysRevA.86.063609},
    url = {https://link.aps.org/doi/10.1103/PhysRevA.86.063609}
}

@misc{supplemental_material,
    title = {{S}upplemental {M}aterials to {V}irtual defect mediated quantum melting in supersolids.},
    author = {L. Tsaloukidis and M. Grossklags and L. Salasnich and A. Mendoza-Coto},
    year = {2026}
}

@article{pretko2018,
    title = {Symmetry-Enriched Fracton Phases from Supersolid Duality},
    author = {Pretko, Michael and Radzihovsky, Leo},
    journal = {Phys. Rev. Lett.},
    volume = {121},
    issue = {23},
    pages = {235301},
    numpages = {7},
    year = {2018},
    month = {Dec},
    publisher = {American Physical Society},
    doi = {10.1103/PhysRevLett.121.235301},
    url = {https://link.aps.org/doi/10.1103/PhysRevLett.121.235301}
}
\onecolumngrid
\section*{End Matter}
\twocolumngrid
\textit{Critical properties of the RG flow.}---The critical behavior resulting from Eq.~\eqref{rg_flow_equations} can be obtained directly from the structure of its RG trajectories close to the boundary separating the weak- and strong-coupling regimes. Introducing $s=\sqrt{d}+\sqrt{e}$, we write $dk_{u}/dl=\beta(a)k_{u}$, with $\beta(a)=2-\pi s/\left(\sqrt{a}\sqrt{s^{2}a+c}\right)$. Along the fixed line $k_{u}=0$, the cosine locking perturbation becomes marginal at $\beta(a_{c})=0$, which gives
\begin{equation}
    a_{c}=\frac{-c+\sqrt{c^{2}+\pi^{2}s^{4}}}{2s^{2}}.
\end{equation}
The RG flow possesses the exact first integral
\begin{equation}
    I(k_{u},a)=k^{2}_{u}-\frac{4a}{\pi^{3}}+\frac{4}{\pi^{2}}\operatorname{arcsinh}\left(\frac{s\sqrt{a}}{\sqrt{c}}\right).
\end{equation}
The critical separatrix is determined by $I=I_{c}\equiv I(0,a_{c})$. Since $\partial_{a}I=-2\beta(a)/\pi^{3}$, an expansion around the marginal fixed point up to quadratic order yields
\begin{equation}
    \delta\equiv{I}-{I}_{c}\simeq k^{2}_{u}-\frac{\beta'(a_{c})}{\pi^{3}}(a-a_{c})^{2}.
\end{equation}
Thus, the trajectories close to criticality have the hyperbolic invariant characteristic of the KT flow, with $\delta=0$ defining the separatrix. On the strong-coupling side, $\delta>0$, the previous result can be inserted into $da/dl=\pi^{3}k^{2}_{u}$, yielding
\begin{equation}
    \frac{d(a-a_{c})}{dl}\simeq\pi^{3}\delta+\beta'(a_{c})(a-a_{c})^{2}.
\end{equation}
Direct integration then lead us to 
\begin{equation}
    l-l_{0}=\frac{1}{\sqrt{\pi^{3}\beta'(a_{c})\delta}}\left[\tan^{-1}\left((a-a_{c})\sqrt{\frac{\beta'(a_{c})}{\pi^{3}\delta}}\right)\right]_{a_{0}}^{a},
\end{equation}
showing that the RG scale at which the flow reaches strong coupling diverges as $l^{\ast}\propto\delta^{-1/2}$. Since $\xi\simeq\xi_{0}e^{l^{\ast}}$, the correlation length therefore displays the KT essential singularity $\xi\sim\xi_{0}\exp(b/\sqrt{\delta})$, where $b>0$ is non-universal.

Moreover, the bare RG couplings depend smoothly on the microscopic parameters used to explore the phase diagram of the system. Denoting these parameters collectively by $\boldsymbol{\mathrm{p}}$, the critical line is defined by $\delta(\boldsymbol{\mathrm{p}})=0$. Accordingly, around a regular point $\boldsymbol{\mathrm{p}}_{c}$ of this line, the distance from criticality varies to leading order as $\delta(\boldsymbol{\mathrm{p}})=\nabla_{\boldsymbol{\mathrm{p}}}\delta|_{c}\cdot(\boldsymbol{\mathrm{p}}-\boldsymbol{\mathrm{p}}_{c})+\cdots$. Hence, in Fig.~\ref{fig1}(b), if we approach the critical line varying density at a fixed interaction strength, we have $\delta\propto n-n_{c}$, close to the transition. Consequently, on the supersolid side, the correlation length exhibits the characteristic behavior $\xi\sim\xi_{0}\exp[b/\sqrt{n-n_{c}}]$, where $b$ is a non-universal parameter.\\

\textit{Positional correlation within the superfluid phase.}---Equation~\eqref{rg_flow_equations} allows us to determine the asymptotic behavior of the positional correlations in the rough phase. Indeed, the flow of the locking coupling can be written as $dk_{u}/dl=(2-\Delta_{q}(a))k_{u}$, with 
\begin{equation}
    \Delta_{q}(a)=\frac{\pi}{a}\frac{s}{\sqrt{s^{2}+c/a}}.
\end{equation}
Here, $\Delta_{q}(a)$ corresponds to the scaling dimension of the vertex operator $e^{iqu}$, and therefore also of the locking operator $\cos(qu)$.

On the superfluid side of the transition, $k_{u}(l)\rightarrow0$ and $a(l)\rightarrow a_{\infty}<\infty$, the infrared theory approaches a point of the Gaussian fixed line. The positional correlation can then be written as
\begin{equation}
    C_{q}(\boldsymbol{r})\equiv\left\langle e^{iq[u(\boldsymbol{r})-u(0)]}\right\rangle=\exp\left[-\frac{q^{2}}{2}\left\langle[u(\boldsymbol{r})-u(0)]^{2}\right\rangle\right],
\end{equation}
where the last equality follows from the Gaussian character of the infrared fluctuations. Under an infinitesimal RG step, a single vertex $e^{iqu}$ acquires the factor $\exp[-\Delta_{q}(a)\,dl]$. Since $C_{q}$ contains two vertex operators, its logarithmic derivative therefore obeys
\begin{equation}
    \frac{d\ln C_{q}}{dl}=-2\Delta_{q}[a(l)].
\end{equation}
At sufficiently large scales, $a(l)\rightarrow a_{\infty}$, and hence $\Delta_{q}[a(l)]\rightarrow\Delta_{q}(a_{\infty})$ becomes constant. Integrating up to the scale $l\simeq\ln(r/\xi_{0})$ gives
\begin{equation}
    C_{q}(r)\sim\left(\frac{\xi_{0}}{r}\right)^{2\Delta_{q}(a_{\infty})}\equiv\left(\frac{\xi_{0}}{r}\right)^{\eta},
\end{equation}
with
\begin{equation}
    \eta=2\Delta_{q}(a_{\infty})=\frac{2\pi}{a_{\infty}}\frac{s}{\sqrt{s^{2}+c/a_{\infty}}}.
\end{equation}
Restoring the original space and imaginary-time coordinates, $r=(x^{2}+A\tau^{2}/B)^{1/2}$ we recover the positional correlation presented in the main text. Finally, we can notice that as the transition is approached from the superfluid side, $a_{\infty}\rightarrow a_{c}$. Since the external potential contribution is marginal at the critical point, $\beta(a_{c})=0$, and therefore $\Delta_{q}(a_{c})=2$. It follows directly that $\eta_{c}=2\Delta_{q}(a_{c})=4$, as stated in the main text.\\

\textit{Perturbative behavior of the renormalized Fourier amplitudes.}---The leading order effect of the positional fluctuations on the density modulation can be obtained directly from the renormalization of the vertex operators associated with the displacement field. Writing the instantaneous density profile as
\begin{equation}
    \rho(x,u)=\sum_{j\geq0}c_{j,\mathrm{MF}}\cos(jq(x-u)),
\end{equation}
and taking into account the symmetry of the fluctuations around a minimum of the locking potential, we can conclude that
\begin{equation}
    \langle\rho(x,u)\rangle=\sum_{j\geq0}\langle c_{j,\mathrm{MF}}\rangle\cos(jqx),
\end{equation}
where $\langle c_{j}\rangle=c_{j,\mathrm{MF}}\langle\cos(jqu)\rangle $. Within the momentum-shell RG, we decompose $u=u_{<}+u_{>}$ and integrate the fast modes $u_{>}$ over an infinitesimal shell. To leading order, the Gaussian average of the vertex operator gives
\begin{equation}
    \left\langle e^{ijqu_{>}}\right\rangle_{>}=\exp\left[-\frac{j^{2}q^{2}}{2}\langle u^{2}_{>}\rangle\right],
\end{equation}
while the running expectation value satisfies
\begin{equation}
    \frac{d}{dl}\ln\langle\cos(jqu)\rangle=-j^{2}\Delta_{q}(l),
\end{equation}
where $\Delta_{q}(l)=2-\beta(a(l))$ is the running scaling dimension of the fundamental vertex $e^{iqu}$. Inside the pinned phase, the perturbative RG remains valid up to the scale $l^{\ast}$ at which the locking coupling becomes of order unity. The contribution of the modes integrated up to this scale therefore yields 
\begin{equation}\label{idcos}
    \langle\cos(jqu)\rangle\simeq\mathcal{B}_{j}\exp\left[-j^{2}\int_{0}^{l^{\ast}}dl\,\Delta_{q}(l)\right],
\end{equation}
where $\mathcal{B}_{j}$ is a non-universal matching factor accounting for the subsequent strong-coupling regime. Close to the critical line, the flow spends an increasingly long RG time in the vicinity of the marginal fixed point, where $\Delta_{q}(l)\simeq\Delta_{q}(a_{c})=2$. Consequently, the integral above is proportional to $l^{\ast}$, considering then that $l^{\ast}\propto\delta^{-1/2}$ we reach to a scaling of the form $\langle\cos(jqu)\rangle\sim\exp\left[-j^{2} b_{\rho}/\sqrt{\delta}\right]$, where $b_{\rho}>0$ is a non-universal function of the microscopic parameter. Hence, as we approach the critical line varying density, for a given two-body interaction, we obtain to leading order
\begin{equation}
    \langle c_{j}\rangle\sim c_{j,\mathrm{MF}}\exp\left[-\frac{j^{2} G}{\sqrt{n-n_{c}}}\right].
\end{equation}
Furthermore, Eq.~\eqref{idcos}, together with the RG trajectories obtained from the numerical integration of Eq.~\eqref{rg_flow_equations}, allows us to evaluate $\langle\cos(jqu)\rangle$ beyond its asymptotic critical form. Retaining the full dependence of $\Delta_{q}[a(l)]$ along the RG flow and taking $\mathcal{B}_{j}=1$, thereby neglecting additional renormalization after the flow enters the strong-coupling regime, we numerically determine the renormalized Fourier amplitudes and reconstruct the corresponding average density profile. The latter is then employed to evaluate the leading renormalization of the superfluid fraction presented in the main text.\\

\textit{Superfluid fraction.}---For a $1$D condensate of $N$ particles of mass $m$ in a system of length $L$ with periodic boundary conditions, the mean-field ground-state is described by $\varphi_{0}(x)=\sqrt{\rho_{0}(x)}e^{i\theta_{0}}$, where $\rho_{0}(x)>0$ is the equilibrium density profile and $\theta_{0}$ is spatially uniform. The superfluid fraction can be defined from the energy cost associated with imposing a small phase twist $\Theta$ across the system, $\Delta E(\Theta)=N\hbar^{2} f_{s}\Theta^{2}/(2mL^{2})+\mathcal{O}(\Theta^{4})$. At the mean-field level, we therefore impose a phase profile $\theta(x)$ satisfying $\theta(L)-\theta(0)=\Theta$, while keeping $\rho_{0}(x)$ fixed. The corresponding energy cost is $\Delta E[\theta]=(\hbar^{2}/2m)\int_{0}^{L}dx\,\rho_{0}(x)\big[\partial_{x}\theta(x)\big]^{2}$. For a fixed $\Theta$, minimization with respect to $\theta(x)$ gives $\partial_{x}\theta(x)=\Theta\big[\rho_{0}(x)\int_{0}^{L}dx'\rho^{-1}_{0}(x')\big]^{-1}$ and, by extension, $f_{s,\mathrm{MF}}=\big[L^{-1}\int_{0}^{L}dx\,n\rho^{-1}_{0}(x)\big]^{-1}$, where $n=N/L$. This is the well-known result for the superfluid fraction of a $1$D supersolid~\cite{platt2024}.

We now consider how this static phase-twist energy is modified by fluctuations beyond mean field. Writing $\psi(x,\tau)=\sqrt{\rho(x,\tau)}e^{i\theta(x,\tau)}$, the phase-dependent terms of the action take the form $\mathcal{S}=\int d\tau dx\big[i\rho\partial_{\tau}\theta+\hbar^{2}\rho(\partial_{x}\theta)^{2}/2m\big]$. To probe the static response, we impose $\theta=\theta(x)$, so that $\partial_{\tau}\theta=0$, and denote by $\Phi$ all remaining fluctuation fields. The Berry term then vanishes and, at the level of approximation retained here, the action reduces to $\mathcal{S}[\theta,\Phi]=(\hbar^{2}/2m)\int d\tau dx\,\rho(x,\tau)\big[\partial_{x}\theta(x)\big]^{2}$. Defining $\langle\cdot\rangle_{0}$ as the average in the absence of the imposed phase twist, the phase-only effective action satisfies $\mathcal{S}_{\mathrm{eff}}[\theta]-\mathcal{S}_{\mathrm{eff}}[0]=-\log\langle e^{-\mathcal{S}[\theta,\Phi]}\rangle_{0}$. Since $\mathcal{S}$ is already second order in $\partial_{x}\theta$, only its first cumulant contributes at $\mathcal{O}(\Theta^{2})$, yielding $\mathcal{S}^{(2)}_{\mathrm{eff}}[\theta]=(\hbar^{2}/2m)\int d\tau dx\,\langle\rho(x,\tau)\rangle_{0}\big[\partial_{x}\theta(x)\big]^{2}$.

For an equilibrium state, time-translation invariance implies that $\langle\rho(x,\tau)\rangle_{0}$ is a $\tau$ independent function, therefore, the static energy cost is $\Delta E_{\mathrm{eff}}[\theta]\simeq(\hbar^{2}/2m)\int_{0}^{L} dx\,\langle\rho(x)\rangle_{0}\big[\partial_{x}\theta(x)\big]^{2}$, which necessarily lead us to conclude that 
\begin{equation}
    \langle f_{s}\rangle\simeq\left[\frac{1}{L}\int_{0}^{L}dx\,\frac{n}{\langle\rho(x)\rangle_{0}}\right]^{-1}.
\end{equation}
In summary, to the order retained here, the effect of fluctuations on the static superfluid response amounts to replacing the mean-field density profile $\rho_{0}(x)$ by the renormalized profile $\langle\rho(x)\rangle_{0}$ in the corresponding phase-twist expression, justifying the prescription employed in the main text.
\newpage
\phantom{a}
\newpage
\setcounter{figure}{0}
\setcounter{equation}{0}

\renewcommand{\thetable}{S\arabic{table}}
\renewcommand{\thefigure}{S\arabic{figure}}
\renewcommand{\theequation}{S\arabic{equation}}
\renewcommand{\thepage}{S\arabic{page}}

\renewcommand{\thesection}{S\arabic{section}}

\onecolumngrid

\begin{center}
{\large \bf Supplemental Material:\\
Virtual defect mediated quantum melting in supersolids}\\
\setcounter{page}{1}
\end{center}

\section{S1. \hspace{0.4mm} Dimensional reduction}
We consider a zero temperature quantum gas of $N$ particles of mass $m$ strongly confined by a harmonic potential along the $y$ and $z$ directions, while remaining free to move along the $x$-direction in the presence of a shallow periodic potential $U(x)=U_{0}\sin^{2}(qx/2)$. The particles interact through an isotropic two-body soft core potential $V(r)$ that decreases monotonically with $r$ and whose Fourier transform develops a negative sector, $\hat{V}(k)<0$. Such interactions are known to favor clustering above a sufficiently large density~\cite{cinti2014,shen2026}.

The condensate wavefunction $\Psi(\boldsymbol{\mathrm{r}})$ is normalized according to $\int d^{3}r\,\lvert\Psi(\boldsymbol{\mathrm{r}})\rvert^{2}=N$. At the mean-field level, the Hamiltonian functional is
\begin{equation}\label{eq:H3D}
    \mathcal{H}[\Psi,\Psi^{\ast}]=\int d^{3}r\left[\frac{\hbar^{2}}{2m}\lvert\boldsymbol{\nabla}\Psi(\boldsymbol{\mathrm{r}})\rvert^{2}+U_{T}(\boldsymbol{\mathrm{r}})\lvert\Psi(\boldsymbol{\mathrm{r}})\rvert^{2}\right]+\frac{1}{2}\int d^{3}r\,d^{3}r'\,V(\lvert\boldsymbol{\mathrm{r}}-\boldsymbol{\mathrm{r}}'\rvert)\lvert\Psi(\boldsymbol{\mathrm{r}})\rvert^{2}\lvert\Psi(\boldsymbol{\mathrm{r}}')\rvert^{2},
\end{equation}
where $U_{T}(\boldsymbol{\mathrm{r}})=m(\omega^{2}_{y}y^{2}+\omega^{2}_{z}z^{2})/2+U(x)$. In the strong-confinement regime, transverse excitations are frozen and the condensate wavefunction can be factorized as $\Psi(\boldsymbol{\mathrm{r}})=\psi(x)\phi(y,z)$, with $\phi(y,z)=\sqrt{2/(\pi\sigma_{y}\sigma_{z})}\exp[-y^{2}/\sigma^{2}_{y}-z^{2}/\sigma^{2}_{z}]$ and $\sigma_{y,z}=\sqrt{2\hbar/(m\omega_{y,z})}$. Since $\int dy\,dz\,\lvert\phi(y,z)\rvert^{2}=1$, the longitudinal field satisfies $\int dx\,\lvert\psi(x)\rvert^{2}=N$.

Integrating Eq.~\eqref{eq:H3D} over the transverse directions yields
\begin{equation}\label{Heff}
    \begin{split}
        \mathcal{H}[\psi,\psi^{\ast}]=\int dx\bigg[&\frac{n\hbar^{2}}{2m}\left(\frac{1}{\sigma^{2}_{y}}+\frac{1}{\sigma^{2}_{z}}\right)+\frac{nm}{8}\left(\omega^{2}_{y}\sigma^{2}_{y}+\omega^{2}_{z}\sigma^{2}_{z}\right)+\frac{\hbar^{2}}{2m}\lvert\partial_{x}\psi(x)\rvert^{2}+U(x)\lvert\psi(x)\rvert^{2}\\
        &+\frac{1}{2}\int dx'\,V_{\mathrm{eff}}(x-x')\lvert\psi(x)\rvert^{2}\lvert\psi(x')\rvert^{2}\bigg],
    \end{split}
\end{equation}
where $n=N/L$ is the linear particle density. Using $\sigma^{2}_{y,z}=2\hbar/(m\omega_{y,z})$, the first two terms in Eq.~\eqref{Heff} integrate to $N\hbar(\omega_{y}+\omega_{z})/2$. Thus, they represent the transverse zero-point energy, which is independent of the longitudinal configuration and can be discarded as an additive constant.

The remaining effect of the confined directions is contained in the effective interaction
\begin{equation}\label{eq:Veff_reduction}
    V_{\mathrm{eff}}(x)=\int dy\,dy'\,dz\,dz'\,V\big(\sqrt{x^{2}+(y-y')^{2}+(z-z')^{2}}\big)\lvert\phi(y,z)\rvert^{2}\lvert\phi(y',z')\rvert^{2}.
\end{equation}
Therefore, $V_{\mathrm{eff}}(x)$ corresponds to the three-dimensional interaction averaged over the transverse probability distributions of the two particles. Its precise form depends on both $V(r)$ and the confinement widths. In the tight-confinement regime $\sigma_{y,z}\ll R_{c}$, where $R_{c}$ denotes the characteristic range of the soft-core, the transverse probability distribution is localized on a scale much smaller than that over which the interaction varies. Consequently, $V_{\mathrm{eff}}(x)\simeq V(\lvert x\rvert)$ to leading order, and the relevant soft-core character and clustering tendency of the interaction are preserved by the dimensional reduction.

Since our aim is to describe the generic behavior of soft-core supersolids in a quasi-one-dimensional geometry, we model the resulting interaction by $V_{\mathrm{eff}}(x)=v_{0}/\big(1+(x/R_{c})^{6}\big)$. This form provides a simple and well known soft-core potential developing clusterization at large enough densities. After dropping the transverse zero-point energy, the last three terms of Eq.~\eqref{Heff} constitute the effective quasi-one-dimensional Hamiltonian presented in Eq.~$(1)$ of the main text.

\section{S2. \hspace{0.4mm }Quadratic Lagrangian density and elastic parameters}\label{section:elastic}
To account for low-energy excitations, we develop a hydrodynamic description of the supersolid phase~\cite{platt2024,mendoza2025}, providing a microscopic derivation of Eq.~$(2)$ of the main text. We start from the grand-canonical partition function $\mathcal{Z}=\int D\psi D\psi^{\ast}\exp[-\mathcal{S}[\psi,\psi^{\ast}]]$, with zero temperature action
\begin{equation}\label{original_action}
    \mathcal{S}[\psi,\psi^{\ast}]=\int_{0}^{\infty}d\tau\int dx~\psi^{\ast}(x,\tau)(\hbar\partial_{\tau}-\mu)\psi(x,\tau)+\int_{0}^{\infty}d\tau~\mathcal{H}[\psi,\psi^{\ast}].
\end{equation}
We consider low-energy excitations around the mean-field ground-state configuration $\psi_{0}(x)$ by allowing fluctuations in the phase condensate and the modulation, together with their corresponding conjugate density fields. An ansatz for the perturbed wavefunction field consistent with the symmetries of the system is~\cite{mendoza2025,mendoza2026}
\begin{equation}\label{ansatz}
    \psi(x,\tau)=\frac{\sqrt{n}}{2\Xi}\sum_{j}c_{j}\left(1+\frac{\delta n(x,\tau)}{n}+\frac{jk_{0}\Pi(x,\tau)}{Z}\right)^{1/2}e^{i[jk_{0}(x+u(x,\tau))+\theta(x,\tau)]}.
\end{equation}
Here, $u(x,\tau)$ is the phonon field associated with translations of the density modulation, $\theta(x,\tau)$ is the condensate phase, and $\Pi(x,\tau)$ and $\delta n(x,\tau)$ are their respective conjugate density fields. Expanding the action to quadratic order in the fluctuation fields, while retaining the full periodic dependence on $u(x,\tau)$ generated by the external lattice, yields
\begin{equation}\label{effective_action}
    \begin{split}
        \mathcal{S}^{(2)}=&\int d\tau dx\bigg\{i\hbar\delta n\,\partial_{\tau}\theta+i\hbar\Pi\,\partial_{\tau}u+\frac{1}{2}\bigg[\frac{n\hbar^{2}}{m}(\partial_{x}\theta)^{2}+\alpha_{nn}(\delta n)^{2}+\alpha_{\pi\pi}\Pi^{2}+\frac{2\hbar^{2}}{m}\Pi\partial_{x}\theta\\
        &+2\alpha_{nu}\delta n\,\partial_{x}u+\alpha_{uu}(\partial_{x}u)^{2}\bigg]-\kappa_{u}\cos(qu)-\kappa_{n}\cos(qu)\delta n-\kappa_{\pi}\sin(qu)\Pi-\kappa_{\pi2}\cos(qu)\Pi^{2}\bigg\}.
    \end{split}
\end{equation}
The periodic contributions in the action which explicitly breaks continuous translational invariance in the system are a direct consequence of the presence of the interaction with the external lattice. In the commensurate regime considered in the main text, the mean-field modulation wave vector is locked to the lattice wave vector, $k_{0}=q$.

Our goal now is to reduce Eq.~\eqref{effective_action} to an effective action for the deformation field of the supersolid $u(x,\tau)$ that allow us to study the supersolid phase transition in the system. We proceed in two steps. First, we eliminate the auxiliary field $\Pi(x,\tau)$ and express the resulting couplings in terms of static responses of the microscopic Hamiltonian. We then integrate out the density and condensate-phase fluctuations, $\delta n(x,\tau)$ and $\theta(x,\tau)$, which enter quadratically at the order considered, to obtain the effective action governing the positional fluctuations of the supersolid. To eliminate the auxiliary field $\Pi(x,\tau)$, which enters the action algebraically, we use the corresponding Euler--Lagrange equation
\begin{equation}\label{EoM1}
    \left[\alpha_{\pi\pi}-2\kappa_{\pi2}\cos(qu)\right]\Pi+i\hbar\partial_{\tau}u+\frac{\hbar^{2}}{m}\partial_{x}\theta-\kappa_{\pi}\sin(qu)=0.
\end{equation}
From here we can conclude that
\begin{equation}\label{reduced_action_pi}
    \begin{split}
        \mathcal{S}^{(2)}=\int dx\,d\tau\bigg[&i\hbar\delta n\,\partial_{\tau}\theta+\frac{n\hbar^{2}}{2m}(\partial_{x}\theta)^{2}+\frac{\alpha_{nn}}{2}(\delta n)^{2}+\alpha_{nu}\delta n\,\partial_{x}u+\frac{\alpha_{uu}}{2}(\partial_{x}u)^{2}-\kappa_{u}\cos(qu)-\kappa_{n}\cos(qu)\delta n\\
        &-\frac{1}{2}\frac{\left[i\hbar\partial_{\tau}u+(\hbar^{2}/m)\partial_{x}\theta-\kappa_{\pi}\sin(qu)\right]^{2}}{\alpha_{\pi\pi}-2\kappa_{\pi2}\cos(qu)}\bigg].
    \end{split}
\end{equation}
In the long-wavelength regime, we should keep the leading external lattice operator $\kappa_{u}\cos(qu)$ and the leading Gaussian gradient terms. The remaining lattice-dependent couplings generate derivative-dressed locking terms, lattice-dependent corrections to the Gaussian coefficients, and higher harmonics such as $\cos(2qu)$. These contributions are sub-leading in the gradient and weak-lattice expansion and will be neglected here. The resulting action is then
\begin{equation}\label{leading_hydro_action}
    \begin{split}
        \mathcal{S}^{(2)}[\theta,\delta n,u]=\int dx\,d\tau\bigg[&i\hbar\delta n\,\partial_{\tau}\theta+\frac{n\hbar^{2}}{2m}(\partial_{x}\theta)^{2}+\frac{\alpha_{nn}}{2}(\delta n)^{2}+\alpha_{nu}\delta n\,\partial_{x}u+\frac{\alpha_{uu}}{2}(\partial_{x}u)^{2}\\
        &-\kappa_{u}\cos(qu)+\frac{\hbar^{2}}{2\alpha_{\pi\pi}}\left(\partial_{\tau}u-\frac{i\hbar}{m}\partial_{x}\theta\right)^{2}\bigg],
    \end{split}
\end{equation}
where we note that for $\kappa_{u}=0$, this action has the standard hydrodynamic structure obtained for one-dimensional supersolids from symmetry arguments~\cite{yoo2010,platt2024}.

The couplings entering Eq.~\eqref{leading_hydro_action} can be determined directly from the energy cost of suitable static deformations of the mean-field ground-state. To extract $\alpha_{uu}$, we consider a slowly varying displacement field whose strain $\partial_{x}u$ alternates between $\pm\epsilon$ over domains of size $\Delta x$, chosen such that $q\epsilon\Delta x\ll1$. This construction probes the elastic response while suppressing the direct pinning contribution of the external lattice. Locally, the modulation wave vector becomes $q(1\pm\epsilon)$ and therefore, $\alpha_{uu}=\left.\partial^{2}_{\epsilon}(\Delta E_{uu}/L)\right|_{\epsilon=0}$, with
\begin{equation}
    \frac{\Delta E_{uu}}{L}=\frac{1}{2}\frac{\mathcal{H}(q+\epsilon q,q,n,0)}{L}+\frac{1}{2}\frac{\mathcal{H}(q-\epsilon q,q,n,0)}{L}-\frac{\mathcal{H}(q,q,n,0)}{L}.
\end{equation}
Here, the first argument of $\mathcal{H}(q_{1},q_{2},n,u_{0})$ specifies the modulation wave vector employed in the energy calculation of the translationally invariant part of the Hamiltonian, while the second specifies the supersolid wave vector used in evaluating the interaction energy with the external lattice. The remaining arguments denote the mean density $n$ and the overall displacement $u_{0}$ relative to the lattice. In each evaluation, the Fourier amplitudes $\{c_{j}\}$ are minimized at fixed $q_{1}$, $q_{2}$, $n$, and $u_{0}$. The locking amplitude $\kappa_{u}$ then follows from the energy cost of a uniform displacement $u(x)=u_{0}$, $\kappa_{u}=q^{-2}\left.\partial_{u_{0}}^{2}(\Delta E_{\kappa}/L)\right|_{u_{0}=0}$, where
\begin{equation}
    \frac{\Delta E_{\kappa}}{L}=\frac{\mathcal{H}(q,q,n,u_{0})}{L}-\frac{\mathcal{H}(q,q,n,0)}{L}.
\end{equation}
Similarly, the density--strain coupling is $\alpha_{nu}=-\left.\partial_{n}\partial_{\epsilon}(\Delta E_{u}/L)\right|_{n=n_{0},\epsilon=0}$, with
\begin{equation}
    \frac{\Delta E_{u}}{L}=\frac{\mathcal{H}(q+\epsilon q,q,n,0)}{L}-\frac{\mathcal{H}(q,q,n,0)}{L},
\end{equation}
while the inverse compressibility coefficient is $\alpha_{nn}=\left.\partial^{2}_{n}(\Delta E_{nn}/L)\right|_{n=n_{0}}$, where
\begin{equation}
    \frac{\Delta E_{nn}}{L}=\frac{\mathcal{H}(q,q,n,0)}{L}-\frac{\mathcal{H}(q,q,n_{0},0)}{L}.
\end{equation}
The remaining coefficient $\alpha_{\pi\pi}$ has a direct hydrodynamic interpretation. From Eq.~\eqref{leading_hydro_action}, the bare kinetic term of the lattice displacement is $\hbar^{2}(\partial_{\tau}u)^{2}/(2\alpha_{\pi\pi})$. Comparing this expression with the kinetic energy density $mn_{n}(\partial_{\tau}u)^{2}/2$ of the normal component, we can conclude that
\begin{equation}\label{alphapipi_normal}
    \alpha_{\pi\pi}=\frac{\hbar^{2}}{mn_{n}},
\end{equation}
where $n_{n}$ is the linear normal density. Since $n_{s}+n_{n}=n$ and $n_{s}=nf_{s}$, Eq.~\eqref{alphapipi_normal} also shows that the coefficient of $(\partial_{x}\theta)^{2}$ in Eq.~\eqref{leading_hydro_action}, after expanding the last term, is $(\hbar^{2}n_{s}/2m)$, as required by the superfluid phase stiffness.

Since our main interest is the positional melting of the supersolid, we now integrate out $\delta n$ and $\theta$. This step does not neglect their fluctuations: both fields enter quadratically at the retained order, and their effects remain encoded in the resulting kernel for $u$. This process yields the positional effective action
\begin{equation}\label{reduced_action4}
    \mathcal{S}_{\mathrm{eff}}[u]=\frac{1}{2}\int\frac{dk\,d\omega}{(2\pi)^{2}}\left[Ak^{2}+B\omega^{2}+\frac{Ck^{2}\omega^{2}}{Dk^{2}+E\omega^{2}}\right]\hat{u}(k,\omega)\hat{u}(-k,-\omega)-\int d\tau dx\,\kappa_{u}\cos(qu),
\end{equation}
with
\begin{equation}
    A=\alpha_{uu}-\frac{\alpha^{2}_{nu}}{\alpha_{nn}},\hspace{0.3cm}B=\frac{\hbar^2}{\alpha_{\pi\pi}},\hspace{0.3cm}C=\left(\alpha_{nn}\frac{\hbar^{2}}{m}+\alpha_{\pi\pi}\alpha_{nu}\right)^{2},\hspace{0.3cm}D=\frac{\alpha^{2}_{nn}\alpha_{\pi\pi}}{m}\left(\alpha_{\pi\pi}n-\frac{\hbar^{2}}{m}\right),\hspace{0.3cm}E=\alpha_{nn}\alpha^{2}_{\pi\pi}.
\end{equation}

The action in Eq.~\eqref{reduced_action4} corresponds to a quantum sine-Gordon model modified by the presence of a nonlocal momentum--frequency interaction. Considering Eq.~\eqref{alphapipi_normal}, we can conclude that $D\propto\big(n-\hbar^{2}/(m\alpha_{\pi\pi})\big)=n-n_{n}=n_{s}$. Consequently, when the superfluid density vanishes, $n_{s}=0$, one has $D=0$ and the non-local contribution reduces to a local $k^{2}$ term, recovering the usual quantum sine-Gordon model. For finite $n_{s}$, instead, the coupling to the gapless condensate phase produces the non-analytic momentum--frequency dependence in Eq.~\eqref{reduced_action4}. Thus, the additional non-local term directly reflects the coexistence of positional order and superfluidity in the supersolid phase.


\section{S3. \hspace{0.4mm} Renormalization Group treatment}\label{rg_treatment_section}
In this section we derive the perturbative Renormalization Group (RG) equations associated with the effective action for the displacement field introduced in Eq.~$(2)$ of the main text,
\begin{equation}\label{eq:RG_start}
    \mathcal{S}_{\mathrm{eff}}[u]=\frac{1}{2}\int\frac{dk\,d\omega}{(2\pi)^{2}}G^{-1}(k,\omega)\lvert\hat u(k,\omega)\rvert^{2}-\kappa_{u}\int dx\,d\tau\cos\big(qu(x,\tau)\big),
\end{equation}
where $G^{-1}(k,\omega)=Ak^{2}+B\omega^{2}+Ck^{2}\omega^{2}/(Dk^{2}+E\omega^{2})$. The coefficients $A,\dots,E$ encode the microscopic properties of the system and are determined by the mean-field hydrodynamic parameters discussed in Sec. \textcolor{blue}{S2}.

To render the local Gaussian sector isotropic while simultaneously measuring space-time coordinates and displacements in units set by the lattice periodicity $2\pi/q$, we introduce $\tilde{x}_{1}=qx/(2\pi)$, $\tilde{x}_{2}=q\sqrt{A/B}\,\tau/(2\pi)$, $\tilde{k}_{1}=2\pi k/q$, $\tilde{k}_{2}=(2\pi/q)\sqrt{B/A}\,\omega$, and $\tilde{u}=qu/(2\pi)$. Dropping the tilde notation hereafter, the resulting dimensionless action reads
\begin{equation}\label{ndact}
    \mathcal{S}_{\mathrm{eff}}[u]=\frac{1}{2}\int_{\lvert\boldsymbol{\mathrm{k}}\rvert<\Lambda}\frac{d^{2}k}{(2\pi)^{2}}\left[ak^{2}_{1}+bk^{2}_{2}+\frac{ck^{2}_{1}k^{2}_{2}}{dk^{2}_{1}+ek^{2}_{2}}\right]\lvert\hat{u}(\boldsymbol{\mathrm{k}})\rvert^{2}-k_{u}\int d^{2}x\cos(2\pi u),
\end{equation}
where the bare couplings are $a(0)=b(0)=\sqrt{AB}(2\pi/q)^{2}$, $c=\sqrt{AB}(2\pi/q)^{2}C$, $d=BD$, $e=EA$, and $k_{u}(0)=\sqrt{B/A}(2\pi/q)^{2}\kappa_{u}$. Since the quadratic kernel is homogeneous of degree two in momentum, all its coefficients are marginal at tree level.

We perform a momentum-shell RG transformation by decomposing the displacement field as $u=u_{<}+u_{>}$, where $u_{>}$ contains the modes within the infinitesimal shell $\Lambda e^{-dl}<\lvert\boldsymbol{\mathrm{k}}\rvert<\Lambda$. Writing $\mathcal{S}_{\mathrm{eff}}=\mathcal{S}_{0}+\mathcal{S}_{\mathrm{int}}$, with $\mathcal{S}_{0}$ and $\mathcal{S}_{\mathrm{int}}$ denoting the quadratic and locking sectors, respectively, the coarse-grained action for the slow field follows from the cumulant expansion $\mathcal{S}'_{\mathrm{eff}}=\mathcal{S}_{0}+\langle\mathcal{S}_{\mathrm{int}}\rangle_{>}-\frac{1}{2}\big(\langle\mathcal{S}_{\mathrm{int}}^{2}\rangle_{>}-\langle\mathcal{S}_{\mathrm{int}}\rangle_{>}^{2}\big)+\mathcal{O}(k^{3}_{u})$.

At first order in $k_{u}$, the Gaussian average over the fast modes gives
\begin{equation}
    \langle\cos\big[2\pi(u_{<}+u_{>})\big]\rangle_{>}=\cos(2\pi u_{<})\exp\big[-2\pi^{2}G_{>}(0)\big],
\end{equation}
where $G_{>}(0)=\int_{\Lambda e^{-dl}<\lvert\boldsymbol{\mathrm{k}}\rvert<\Lambda}d^{2}k\,G(\boldsymbol{\mathrm{k}})/(2\pi)^{2}$. Writing $G^{-1}(\boldsymbol{\mathrm{k}})=k^{2}\mathcal{D}(\varphi)$ in polar coordinates, with $\mathcal{D}(\varphi)=a\cos^{2}\varphi+b\sin^{2}\varphi+c\cos^{2}\varphi\sin^{2}\varphi/(d\cos^{2}\varphi+e\sin^{2}\varphi)$, the radial shell integration gives a factor $dl$, while the angular integration yields $G_{>}(0)=F(a,b)dl/(2\pi^{2})$, where
\begin{equation}\label{eq:F_ab}
    F(a,b)=\frac{1}{2}\int_{0}^{2\pi}\frac{d\varphi}{\mathcal{D}(\varphi)}=\frac{\pi}{\sqrt{ab}}\,\frac{\sqrt{bd}+\sqrt{ae}}{\sqrt{ae+bd+c+2\sqrt{abde}}}.
\end{equation}
Restoring the cutoff contributes the usual factor $e^{2dl}$ from the rescaling of the two-dimensional measure, yielding
\begin{equation}\label{RG1}
    \frac{dk_{u}}{dl}=[2-F(a,b)]k_{u}.
\end{equation}

The quadratic momentum-frequency sector is renormalized at second order in $k_{u}$. Introducing the center-of-mass and relative coordinates $\boldsymbol{\mathrm{R}}=(\boldsymbol{\mathrm{x}}_{1}+\boldsymbol{\mathrm{x}}_{2})/2$ and $\boldsymbol{\mathrm{r}}=\boldsymbol{\mathrm{x}}_{1}-\boldsymbol{\mathrm{x}}_{2}$, the product of two cosine operators separates into a neutral contribution proportional to $\cos\big[2\pi(u_{<}(\boldsymbol{\mathrm{x}}_{1})-u_{<}(\boldsymbol{\mathrm{x}}_{2}))\big]$ and a same-charge contribution generating the higher harmonic $\cos\big[4\pi u_{<}(\boldsymbol{\mathrm{R}})\big]$. The latter is irrelevant near the transition and lies beyond the truncation retained here.

The neutral contribution can be physically understood in terms of tightly bound defect--antidefect pairs, as in the conventional Kosterlitz--Thouless RG treatment. Upon increasing the short-distance cutoff, the smallest pairs are incorporated into the effective elastic response governing long-wavelength fluctuations. Their contribution can be evaluated by integrating over relative separations within an infinitesimal annulus, $1<\lvert\boldsymbol{\mathrm{r}}\rvert<e^{dl}$, where distances are measured in units of the short-distance cutoff~\cite{Kardar2007}.

Since the slow field varies weakly over these distances, we expand $u_{<}(\boldsymbol{\mathrm{R}}+\boldsymbol{\mathrm{r}}/2)-u_{<}(\boldsymbol{\mathrm{R}}-\boldsymbol{\mathrm{r}}/2)=\boldsymbol{\mathrm{r}}\cdot\nabla u_{<}(\boldsymbol{\mathrm{R}})+\mathcal{O}(r^{3})$, so that $\cos(2\pi\boldsymbol{\mathrm{r}}\cdot\nabla u_{<})\simeq1-2\pi^{2}r_{\mu}r_{\nu}\partial_{\mu}u_{<}\partial_{\nu}u_{<}$. Within the leading isotropic short-distance approximation, the angular integration gives
\begin{equation}
    \int_{1<\lvert\boldsymbol{\mathrm{r}}\rvert<e^{dl}}d^{2}r\,r_{\mu}r_{\nu}=\int_{1}^{e^{dl}}dr\,r^{3}\int_{0}^{2\pi}d\varphi\,\hat{r}_{\mu}\hat{r}_{\nu}=\pi\,dl\,\delta_{\mu\nu}+\mathcal{O}(dl^{2}).
\end{equation}
Consequently, the leading correction to the local Gaussian action is $\delta\mathcal{S}_{0}=(\pi^{3}k^{2}_{u}/2)dl\int d^{2}x\big[(\partial_{1}u_{<})^{2}+(\partial_{2}u_{<})^{2}\big]$, yielding
\begin{equation}\label{RG2}
    \frac{da}{dl}=\frac{db}{dl}=\pi^{3}k^{2}_{u}.
\end{equation}

Although the full Gaussian propagator is generally anisotropic, the leading short-distance correction is isotropic within the RG procedure adopted above. Moreover, the local gradient expansion generates no correction with the non-analytic momentum dependence associated with $c$, $d$, and $e$, which therefore remains unchanged at this order. Since the bare theory satisfies $a(0)=b(0)$ and both coefficients obey the same flow equation, the RG trajectory remains on the sub-manifold $a(l)=b(l)$. On this sub-manifold, Eq.~\eqref{eq:F_ab} reduces to
\begin{equation}\label{RG3_result}
    F(a)=\frac{\pi}{a}\,\frac{\sqrt{d}+\sqrt{e}}{\big[(\sqrt{d}+\sqrt{e})^{2}+c/a\big]^{1/2}}.
\end{equation}
Substituting Eq.~\eqref{RG3_result} into Eq.~\eqref{RG1} and Eq.~\eqref{RG2} gives the coupled RG equations presented in Eq.~$(3)$ of the main text.

\section{S4. \hspace{0.4mm} Sine-Gordon soliton loops}
In this section, we determine the minimum action defect loop under the constraint of a fixed enclosed area in the $x_{1}$--$x_{2}$ plane, where the renormalized $x_{1}$ and $x_{2}$ coordinates were defined in Sec. \textcolor{blue}{S3}. Departing from the dimensionless action in Eq.~\eqref{ndact}, we first consider a straight soliton wall with constant slope, whose trajectory is $x_{s,1}(x_{2})=vx_{2}+x_{0}$, with $x_{0}$ representing the initial condition and $v=dx_{1}/dx_{2}$. Introducing the relative coordinate $s=x_{1}-x_{s,1}(x_{2})$, we write the corresponding deformation field as $u(x_{1},x_{2})=f(s)$. Its Fourier transform then yields
\begin{equation}
    \hat{u}(k_{1},k_{2})=\int dx_{1}dx_{2}\,e^{-i(k_{1}x_{1}+k_{2}x_{2})}u(x_{1},x_{2})=2\pi e^{-ik_{1}x_{0}}\hat{f}(k_{1})\delta(k_{2}+vk_{1}),
\end{equation}
where $\hat{f}(k_{1})=\int ds\,e^{-ik_{1}s}f(s)$. Therefore, after a substitution of $\hat{u}(k_{1},k_{2})$ into Eq.~\eqref{ndact}, the action cost of a straight wall extending over an interval $L_{2}$ along $x_{2}$ is given by
\begin{equation}\label{wall}
    \Delta\mathcal{S}_{\mathrm{eff}}=\mathcal{S}_{\mathrm{eff}}[u]-\mathcal{S}_{\mathrm{eff}}[0]=L_{2}\int_{-\infty}^{\infty}ds\bigg[\frac{\Gamma(v)}{2}\big(f'(s)\big)^{2}+k_{u}\big[1-\cos(2\pi f(s))\big]\bigg],
\end{equation}
where $\Gamma(v)=a+bv^{2}+cv^{2}/(d+ev^{2})$.

The optimal soliton profile follows from the Euler--Lagrange equation associated with Eq.~\eqref{wall}, subject to the boundary conditions $f(-\infty)=0$ and $f(+\infty)=1$, which yields
\begin{equation}\label{eq:ELkink}
    \Gamma(v)f''(s)=2\pi k_{u}\sin\big(2\pi f(s)\big).
\end{equation}
Multiplying Eq.~\eqref{eq:ELkink} by $f'(s)$ and using the boundary conditions gives the first integral
\begin{equation}
    \frac{\Gamma(v)}{2}\big[f'(s)\big]^{2}=k_{u}\big[1-\cos\big(2\pi f(s)\big)\big].
\end{equation}
Choosing the center of the wall at $s=0$, we obtain
\begin{equation}
    f_{v}(s)=\frac{2}{\pi}\arctan\left(e^{s/\lambda(v)}\right),
\end{equation}
where $\lambda(v)=(2\pi)^{-1}\sqrt{\Gamma(v)/k_{u}}$ is the soliton width. Substituting this solution into Eq.~\eqref{wall}, the action per unit length of a straight wall becomes $4\sqrt{k_{u}\Gamma(v)}/\pi$. This means that when the local curvature is much larger than the soliton width, the action of a closed loop can be estimated as
\begin{equation}\label{loopaction}
    \mathcal{S}_{\mathrm{loop}}=\frac{4\sqrt{k_{u}}}{\pi}\oint_{\mathcal{C}}\sqrt{\Gamma\big(v(x_{2})\big)}~\lvert dx_{2}\rvert,
\end{equation}
where $v(x_{2})=dx_{1}(x_{2})/dx_{2}$. We now determine the loop shape that minimizes Eq.~\eqref{loopaction} for a fixed enclosed area. Considering reflection symmetry of the optimal loop trajectory with respect to both coordinate axes, it is sufficient to calculate its upper-right quarter. Hence, the total enclosed area is
\begin{equation}
    \mathcal{A}=4\int_{0}^{x_{2,\mathrm{max}}}dx_{2}\,x_{1}(x_{2}).
\end{equation}
Since the overall prefactor in Eq.~\eqref{loopaction} does not affect the optimal trajectory, the constrained minimization can be performed using
\begin{equation}
    \mathcal{S}_{\Lambda}[x_{1}]=\int_{0}^{x_{2,\mathrm{max}}}dx_{2}\left[\sqrt{\Gamma(v)}+\Lambda x_{1}\right],
\end{equation}
where $\Lambda$ is the Lagrange multiplier associated with the fixed-area constraint. The corresponding Euler--Lagrange equation yields the parametric solution
\begin{equation}\label{soliton_loop_eq3}
    \begin{split}
        x_{1}(v)&=-\frac{1}{\Lambda}\bigg[\sqrt{\Gamma(v)}-v\frac{d}{dv}\sqrt{\Gamma(v)}\bigg],\\
        x_{2}(v)&=\frac{1}{\Lambda}\frac{d}{dv}\sqrt{\Gamma(v)}.
    \end{split}
\end{equation}
Equation~\eqref{soliton_loop_eq3} describes the right branch of the optimal loop as $v$ varies from $-\infty$ to $+\infty$. The left branch is obtained by reflection, $x_{1}\rightarrow -x_{1}$. The Lagrange multiplier $\Lambda$ fixes the overall size of the defect loop trajectory and is determined by imposing the prescribed enclosed area $\mathcal{A}$. The procedure described above was used to determine the optimal defect loop shown in Fig.~$1$(c) of the main text, where a large enclosed area was chosen to ensure that the local radius of curvature remains much larger than the soliton width.
\end{document}